# Gravitational waves from orbiting binaries without general relativity: a tutorial

Robert C. Hilborn[a)]

*American Association of Physics Teachers*

*College Park, MD 20740*

**Abstract**

This tutorial leads the reader through the details of calculating the properties of gravitational waves from orbiting binaries, such as two orbiting black holes. Using analogies with electromagnetic radiation, the tutorial presents a calculation that produces the same dependence on the masses of the orbiting objects, the orbital frequency, and the mass separation as does the linear version of General Relativity (GR). However, the calculation yields polarization, angular distributions, and overall power results that differ from those of GR. Nevertheless, the calculation produces waveforms that are nearly identical to the pre-binary-merger portions of the signals observed by the Laser Interferometer Gravitational-Wave Observatory (LIGO-Virgo) collaboration. The tutorial should be easily understandable by students who have taken a standard upper-level undergraduate course in electromagnetism.[a]

## I.   INTRODUCTION

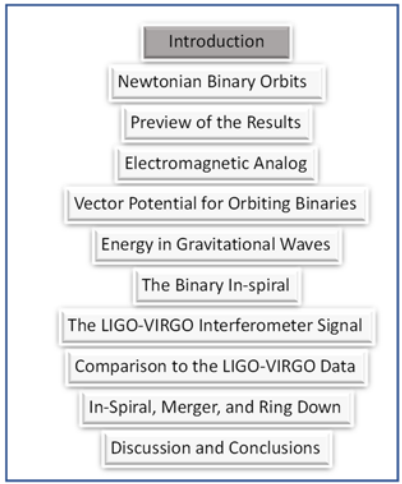

The recent observations[1, 2, 3, 4] of gravitational waves by the Laser Interferometer Gravitational-Wave Observatory (LIGO-Virgo) collaboration have excited the physics community and the general public.[b] The ability to detect gravitational waves (GWs) directly provides a "new spectroscopy" that is likely to have a profound influence on our understanding of the structure and evolution of the universe and its constituents. In addition, the LIGO-Virgo collaboration provides strong evidence that most of the observed GWs can be attributed to the in-spiraling of two mutually orbiting black holes, ending with the coalescence of the black holes,[5] thereby enhancing our confidence that black holes exist. However, to many physicists and certainly to almost all students, the derivations of the expressions to reveal the characteristics of the sources from the LIGO-Virgo observations seem rather challenging due to the complicated formalism of General Relativity (GR). This tutorial provides a derivation of that link based on analogies with electromagnetic waves in a form that most upper-level physics students should be able to understand.

---

[a] Version 2 of this tutorial adds the "integrated orbital phase" method of describing chirp signals and a detailed derivation of the interaction of a vector gravitational wave with a LIGO-Virgo interferometer detector.
[b] A list of all verified gravitational wave observations can be found on Wikipedia: https://en.wikipedia.org/wiki/List_of_gravitational_wave_observations

Most of the popular writing about gravitational waves claims that GR was the first theory to predict gravitational waves. Those claims are certainly false. It is important to realize that almost any relativistic theory of gravity predicts the existence of gravitational waves traveling at a finite speed [Refs. 6, 7, and 8, page 242]. This tutorial provides a specific example of that claim. In fact, there were several attempts to build relativistic theories of gravity before Einstein published his geometrodynamics version (General Relativity) in 1915. Those involved include Heaviside, Poincaré, Nordstrom and even Einstein himself.[c] (For a summary of the history of those attempts, see Refs. 6, 7, and 9.) All those theories predicted the existence of gravitational waves though few of the authors worked out the detailed formalism.

This tutorial treats the relationship between gravitational waves and their sources in analogy with the relationships for electromagnetic (E&M) waves, as a physicist might have done prior to 1915—after special relativity had been introduced but before Einstein formulated GR. It turns out that such a treatment produces most of the GR features of gravitational waves, but not all, as one might expect, since GR is a tensor theory and electromagnetism is essentially a vector theory.

You may know that in GR, the "force" due to gravitation (as expressed for example in Newton's Law of Gravity) is replaced by the curvature of space-time. The curvature in a space-time region is determined by the distribution of mass and energy and is represented mathematically by a tensor. In the "E&M-like" model of gravitational waves developed in this tutorial, we will use mass and mass currents as sources of gravitational waves, just as we use electric charge and electric currents as sources of electromagnetic waves in classical electrodynamics. These waves propagate in the "flat" space-time of special relativity. Just as with standard Newtonian gravitation, in this model masses respond to the local gravitational field $\vec{g}$, which now includes a time-dependent component.

---

[c] Tony Rothman, "The Secret History of Gravitational Waves," Am. Scientist 106, 96-103 (2018).

It is important to note that this E&M-like model is not a complete relativistic theory of gravity.[10, 11] The goals are more modest: understanding the production of gravitational waves from orbiting binaries and the detection of those waves far from their sources. The tutorial is designed is to help you understand the physics behind gravitational waves without needing to learn about metric tensors, Riemannian curvature tensors, and the other mathematical ingredients of GR. I hope you will find it satisfying to be able to understand the basic physics behind gravitational waves and their properties. The results of the calculations presented in this tutorial are in surprisingly good agreement with the LIGO-Virgo observations and the calculations from numerical GR, at least up to the time at which the two orbiting objects "collide."

The arguments presented here should be readily accessible to upper-level undergraduate students. I have included sufficient comments and details to make the entire calculation relatively (no pun intended!) self-contained. Although each of the steps in the calculation is straightforward, there are in fact many steps; so, I urge you to tackle the calculation over a period of several days. You are also encouraged to work with others and to discuss both the mathematical details of the calculations and their conceptual meaning.

GR is inherently a theory that is nonlinear in the space-time metric tensor, the basic building block from which the curvature of space-time is calculated. Almost all analytical GR calculations of the properties of gravitational waves drop the nonlinear terms and keep just the linear ones.[8, 12, 13] The resulting wave equations are very similar to those used in E&M for electromagnetic waves. So, it is not surprising that the E&M-like model used in this tutorial gives results that are similar to those from the linear version of GR (henceforth called "linear GR"). In particular, the E&M-like calculation shows "naturally" that gravitational waves are transverse and for an orbiting binary source, the radiated energy has the dependence on mass, orbital frequency, and mass separation predicted by linear GR. The effects of special relativity are included simply by using gravitational potential functions that depend on the so-called retarded time relative to the observation time. No four-vectors or Lorentz transformations are needed.

> You will find short **Exercises** (highlighted in gray) throughout the tutorial aimed at filling in details of the calculation or extracting insight from the results of various steps along the way. \***Exercises** are particularly important. However, if you are in a hurry to get to the final results, you may skip the exercises without loss of continuity.

> I have also included some **Discussion Questions** aimed to stimulate conversations about basic concepts. Tackle these questions with your friends. **Asides** contain comments about technical details that can be skipped without affecting your understanding of the main points of the tutorial.
>
> In addition to the exercises and discussion questions, the tutorial contains a few "**meta-moments**": suggestions for reflecting on what you experienced, how you drew on what you had learned elsewhere, and what you recognized for the first time.
>
> Since this tutorial is designed for students, the author will greatly appreciate receiving feedback from students, particularly where there may be ambiguities and, heaven-forbid, where there are mistakes. Please send comments to rhilborn@aapt.org.

The strategy of the calculation is the following:

1. Use Newtonian mechanics to describe the orbits of the binary objects and how those orbits decay as the system loses energy.
2. Formulate the gravitational vector potential function due to the orbiting binary objects, taking into account the time delay between events at the binaries and measurements at the observation point.
3. From the gravitational vector potential function, calculate the gravitational radiation field.
4. Employ the gravitation radiation field to find the energy per unit time carried away from the binaries by gravitational waves.
5. Combine the orbital dynamics with the energy loss and the response of a LIGO-Virgo detector to calculate the gravitational waveform that would be seen by LIGO-Virgo.

Section IX compares the waveforms from this calculation with the LIGO-Virgo observations and shows how information about the orbiting black holes is extracted from the observations. Section X provides a phenomenological description of the merger and ring-down of the black holes, while Section XI gives a recap of the tutorial.

The display box at the beginning of each section lets you know where you are in the tutorial.

## II.    NEWTONIAN DYNAMICS FOR THE ORBITING BINARIES

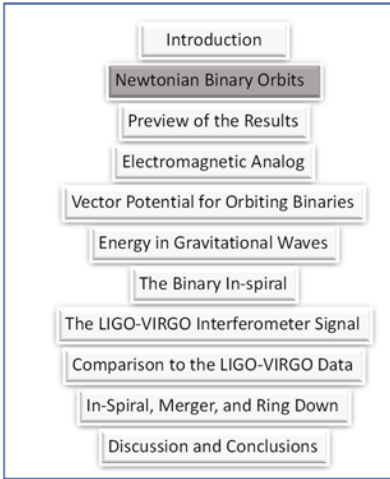

To explain the recent observations of gravitational waves by the LIGO-Virgo team,[1, 2, 3, 4, 14] we will focus attention on the waves emitted by two masses orbiting their common center of mass. The key concept is that the orbiting system emits gravitational waves, which carry energy away from the system, leading to a decay of the orbits, that is, there is an "in-spiral" of the masses.

For simplicity's sake, we will assume the orbits are circular in the absence of gravitational radiation and that when radiation is taken into account, the change in orbital properties is sufficiently slow that a circular orbit is still a reasonable model for the orbital dynamics over sufficiently small time steps. This assumption is violated in the last moments before the objects collide, but nevertheless the circular orbit model provides a surprisingly good approximation.[15]

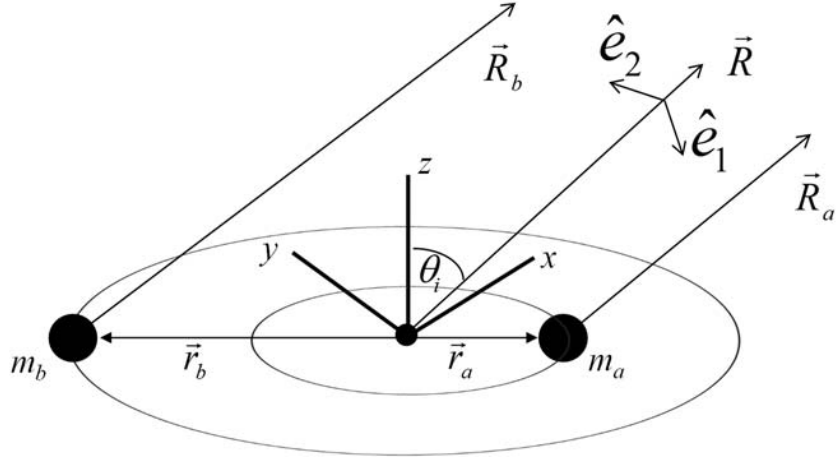

Fig. 1.  A perspective view of the geometry of the binary orbits. The *x-y* plane is chosen to coincide with the plane of the orbits. *z* is perpendicular to that plane.  The distance between the two masses is given by $r = r_a + r_b$ and the vectors $\vec{R}$, $\vec{R}_a$, and $\vec{R}_b$ point from the center of mass (small filled circle), mass $m_a$, and mass $m_b$, respectively, to the (distant) observation point. $\hat{e}_1$ and $\hat{e}_2$ are unit vectors perpendicular to $\vec{R}$ . ( $\hat{e}_1$ and $\hat{e}_2$ are equivalent to the usual spherical coordinate unit vectors $\hat{\theta}$ and $\hat{\phi}$ ). The *x-y* axes are aligned (without loss of generality because of the cylindrical symmetry around the *z* axis) so that $\vec{R} \cdot \hat{y} = 0$. The angle $\theta_i$ between the *z* axis and the observation direction is called the orbital inclination angle in the astrophysics literature.

For the geometry indicated in Fig. 1, the lower-case position vectors $\vec{r}_a$ and $\vec{r}_b$ give the positions of the masses relative to the center of mass, which we may assume to be at rest. (If we are moving relative to the center of mass, there will be, of course, a Doppler shift of the wave frequency.) Recalling that $m_a r_a = m_b r_b$ for distances measured from the center of mass and choosing $\vec{r}$ to be parallel to $\vec{r}_a$, we may write

$$\vec{r}_a = \frac{m_b \vec{r}}{m_a + m_b} \quad \vec{r}_b = -\frac{m_a \vec{r}}{m_a + m_b}. \tag{1}$$

Assuming that Newtonian dynamics applies to the orbits (at least as a first approximation), we recall that Newton's Second Law for mass $m_a$ requires that

$$F_{\text{on } a} = G\frac{m_a m_b}{r^2} = m_a \frac{v_a^2}{r_a}. \tag{2}$$

$G$ is the usual Newtonian gravitational parameter. Invoking Eq. (1) and

$$v_a = \omega r_a, \tag{3}$$

where $\omega$ is the orbital angular frequency (in radians/sec), yields

$$\omega^2 = G\frac{m_a + m_b}{r^3} = \frac{GM}{r^3}, \tag{4}$$

where $M = m_a + m_b$ is the sum of the two masses. Eq. (4) is Kepler's Third Law for the binary system. We will use Eq. (4) many times in what follows.

**Exercise:** Compare Eq. (4) to the formulation of Kepler's Third Law in your favorite introductory physics book. What is different? Show how Eq. (4) reduces to the usual text book formulation under certain assumptions.

**\*Exercise**: This exercise sets up the arguments in the next few paragraphs. Suppose I see a flash of light emitted by a magnetic eruption on the surface of the Sun at precisely 9:21 am Eastern Daylight time. When did that magnetic eruption occur on the Sun? The lesson here is that what we see "now" on Earth occurred earlier on the Sun. For historical reasons, the time at which the event occurred is called the "retarded time." The retarded time concept captures a key principle: it takes some time for "signals" from the Sun to reach the Earth. The generalization of that principle will play a key role in the analysis that follows. In formal terms, we write the retarded time $t_{\text{ret}}$ as $t_{\text{ret}} = t - R/c$, where $t$ is time the signal arrives at the observer, $R$ is the distance from the source to the observer, and $c$ is the signal speed.

Ignoring for the moment any energy loss mechanisms, and introducing the observation time $t$, the speed of light $c$, the coordinate system illustrated in Fig. 1, and the retarded times for each of the two masses,

$$t_{Ra} = t - R_a/c \text{ and } t_{Rb} = t - R_b/c, \tag{5}$$

we may write the position and velocity for mass $m_a$ as

$$\vec{r}_a = r_a \left[ \hat{x}\cos(\omega t_{Ra}) + \hat{y}\sin(\omega t_{Ra}) \right] \tag{6}$$

$$\vec{v}_a = \omega r_a \left[ -\hat{x}\sin(\omega t_{Ra}) + \hat{y}\cos(\omega t_{Ra}) \right]. \tag{7}$$

The position and velocity for $m_b$ are given by analogous expressions:

$$\vec{r}_b = -r_b \left[ \hat{x}\cos(\omega t_{Rb}) + \hat{y}\sin(\omega t_{Rb}) \right] \tag{8}$$

$$\vec{v}_b = \omega r_b \left[ \hat{x}\sin(\omega t_{Rb}) - \hat{y}\cos(\omega t_{Rb}) \right]. \tag{9}$$

For later use, we will need

$$\vec{r} = r \left[ \hat{x}\cos\omega t_R + \hat{y}\sin\omega t_R \right], \tag{10}$$

where $t_R = t - R/c$.

The total mechanical energy for the orbiting binaries is given by

$$U_{total} = \tfrac{1}{2}m_a v_a^2 + \tfrac{1}{2}m_b v_b^2 - G\frac{m_a m_b}{r} = -G\frac{m_a m_b}{2r}, \tag{11}$$

where the last equality follows from the use of Eqs. (3) and (4). As the system radiates gravitational wave energy, the mechanical energy of the binary system decreases (becomes more negative), which means that the distance between the two masses decreases, and by Eq. (4), the orbital frequency increases. It turns out that the energy carried away by a gravitational wave has a strong dependence on the orbital frequency, leading to a large increase in wave energy emitted as the distance between the masses decreases.

> **Meta-moment**: Did any parts of this section surprise you? What aspects of Newtonian mechanics did it make you think carefully about?

## III. PREVIEW OF THE RESULTS

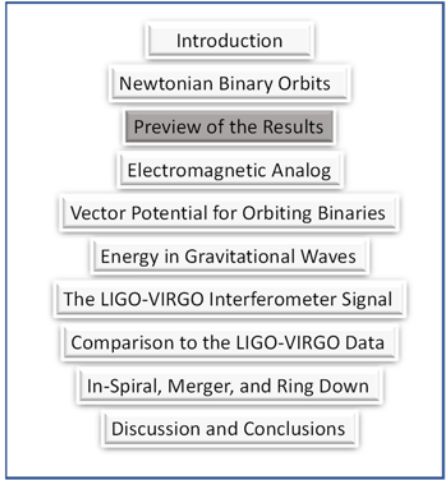

Let's immediately jump to the results of the analysis, which will be presented in detail in the following sections. The critical result is that the orbiting binaries lose energy to gravitational waves. The gravitational wave power emitted (for an otherwise isolated system, equivalent to the rate of orbital energy loss) is given by

$$\frac{dE_G}{dt} = N \frac{G(\eta M)^2 r^4 \omega^6}{c^5}, \qquad (12)$$

where $\eta = m_a m_b / (m_a + m_b)^2$ is a dimensionless ratio of the masses. $N$ is a numerical factor that depends on the details of the theory used to derive the result. For GR, we have $N = 32/5$ while the "E&M-like" calculation in this tutorial gives $N = 2/5$.

---

**Exercise**: Show that $\eta M$ is the standard "reduced mass" used in classical mechanics. Show that for two equal masses $\eta = 1/4$ (its largest value) and that $\eta \approx m_{small} / m_{large}$ when one mass is much larger than the other.

---

As the system loses energy, the distance between the two masses decreases. See Eq. (11). We will later find that the rate ($\dot{r}$) at which that separation $r$ changes is given by a relatively simple expression:

$$\dot{r} = \frac{-2N(GM)^3 \eta}{c^5 r^3}, \qquad (13)$$

**\*Exercise:** Show how Eqs. (11) and (12) lead to Eq. (13). Hint: use Kepler's Third Law to eliminate $\omega$ in favor of $r$.

---

We can repackage the right side of Eq. (13) to uncover some of the physics of the result. First, we introduce $r_s$, the "Schwarzschild radius" associated with a mass. For our purposes, this radius is the distance from the center of a spherically symmetric mass at which the escape speed, calculated from Newtonian mechanics, is equal to the speed of light. $r_s$ is named in honor of Karl Schwarzschild, who in 1915 worked out the solutions to the Einstein equation of General Relativity for the case of a static, spherically symmetric mass, the same year in which Einstein published his eponymous equation. The important point to note is that the Schwarzschild radius sets the distance scale for any relativistic theory of gravity.

---

**\*Exercise**: Recall that the escape speed $v_{esc}$ for a mass $m$ starting at a distance $r_0$ from the center of a spherically symmetric mass $m_0$ satisfies

$$\tfrac{1}{2} m v_{esc}^2 = \frac{G m_0 m}{r_0} \ . \tag{14}$$

Show that the distance $r_s$ for which $v_{esc} = c$ is given by

$$r_s = \frac{2 G m_0}{c^2} \ . \tag{15}$$

---

It is interesting to note that the concept of an object whose gravity is sufficiently strong to prevent light from escaping from its pull was introduced in the 1700s independently by John Michell, an English clergyman and scientist, and Pierre-Simon Laplace, a French nobleman, mathematician and scientist.[16]

Let's now use the sum of the Schwarzschild radii for the two binary masses $r_S \equiv 2GM/c^2$ to repackage Eq. (13) as

$$\dot{r} = \frac{-\eta N c}{4}\left(\frac{r_S}{r}\right)^3 \ . \tag{16}$$

Eq. (16) tells us that for $r \gg r_S$, the magnitude of the rate of change of the separation is very small, but as $r \to r_S$, we have $|\dot{r}| \to \eta 8c/5$ for the GR calculation (for which $N = 32/5$) and $|\dot{r}| \to \eta c/10$ for the E&M-like calculation $(N = 2/5)$. Given Eq. (16), a relatively simple differential equation, you can find how the separation between the two binary masses changes with time. The solution to that differential equation can be written as

$$\begin{aligned} r^4(t) &= r_i^4 - N\eta r_S^3 c\,(t - t_i) \\ &= N\eta r_S^3 c\,(t_c - t) \ . \end{aligned} \tag{17}$$

The last equality in Eq. (17) introduces the coalescence time $t_c$, the time at which the separation between the masses goes to 0 (in this model).

Once we have Eq. (17), it is easy to set up a spreadsheet or a simple computer program to see how the binary orbits evolve with time. A visual simulation of the in-spiraling binaries as a GlowScript (VPython) program is available at http://www.glowscript.org/#/user/rhilborn/folder/Public/program/BinaryInSpiral. (Just click on the hyperlink and the program should open in a browser.)

**Exercise**: Verify the result stated in Eq. (16).
**Exercise**: Verify that Eq. (17) is a solution of Eq. (16).
**Exercise**: Show that the solution to Eq. (16) can also be written as $r^4(t) = \eta N c r_S^3 (t_c - t)$, where $t_c$ is the so-called coalescence time, the time at which the separation between the masses goes to 0 (in our model).

**Discussion Question**: Provide reasons why the maximum rate of approach of the two masses should be about $c$.
**Exercise**: Use Kepler's Third Law to show that the orbital angular frequency of the orbiting binaries can be written as

$$\omega(t) = \frac{c}{\sqrt{2}} \sqrt{\frac{r_S}{r^3(t)}} \tag{18}$$

**Exercise:** Show that in our non-relativistic model, the speed of the binary objects in their circular orbits ($v = \omega r / 2$ for the case of equal masses) can exceed $c$.
**Meta-moment**: This section stated results without a detailed derivation. Were you okay with that? Did the section motivate you for what is to come or was it just frustrating?

## IV. ELECTROMAGNETIC WAVES PROVIDE A PARADIGM FOR GRAVITATIONAL WAVES

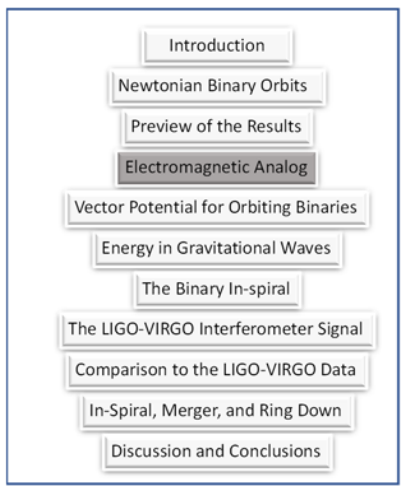

The analogy between electromagnetism (E&M) and gravitation makes use of the equivalent roles of $1/4\pi\varepsilon_0$ and $G$, the Newtonian gravitational parameter, when SI units are used in the expressions relating sources to potentials and fields. In E&M, you should recall that electromagnetic waves are emitted when electric charges accelerate. Similarly, gravitational waves are emitted when masses undergo appropriate forms of acceleration.

First let's review some basic E&M and the description of electromagnetic waves and their relation to potential functions.

If the material looks familiar, please feel free to skip ahead. If your study of E&M has not yet included scalar and vector potentials, particularly as they are applied to E&M waves, don't panic. The following should provide enough explanation to make the results seem plausible.

As is well known (see your favorite E&M text book), electric and magnetic fields can be expressed in terms of two potential functions[17,18]: a scalar potential $\Phi$ and a vector potential $\vec{A}$. The electric ($\vec{E}$) and magnetic ($\vec{B}$) fields are found by calculating the space and time derivatives of the potential functions:

$$\vec{E} = -\vec{\nabla}\Phi - \frac{\partial \vec{A}}{\partial t} \tag{19}$$

$$\vec{B} = \vec{\nabla} \times \vec{A}. \tag{20}$$

> **Exercise.** Write an expression for the scalar potential $\Phi$ associated with a single electric charge so that taking the gradient (spatial derivative) gives Coulomb's Law for the electric field due to a point electric charge.
> **Exercise.** If you have a collection of point charges, how do you find the corresponding scalar potential? Hint: superposition.
> **Exercise.** In your favorite E&M book look up the vector potential due to a time-independent electrical current. See how the magnetic field is calculated from the vector potential. Does a time-independent current produce an electric field? What happens if the current is changing with time?

Now let's take a look at E&M waves. There are several critical issues. First, as mentioned previously, E&M waves are emitted by accelerating charges, so we must deal with time-dependent situations. Second, once the waves are emitted and are far from their sources (far in this case means many, many wavelengths), the oscillating electric and magnetic fields that make up E&M waves are perpendicular to the direction of wave propagation (recall "transverse" waves) and the field amplitudes fall off as $1/R$ where $R$ is the distance from the source to the observation point. (See Fig. 2.)

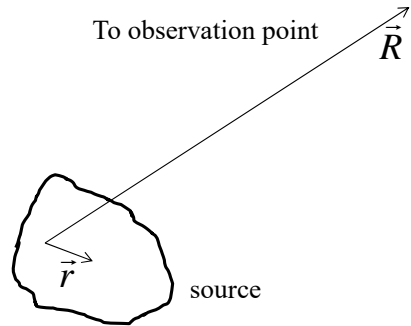

Fig. 2. The vector $\vec{r}$ denotes positions within the source region. $\vec{R}$ is a vector pointing from the coordinate system origin to a (distant) observation point. (Throughout this tutorial, I shall use upper-case ("large") $R$ to denote the distance between the source and a distant observation point and lower-case ("small") $r$ to denote distances within the source.)

For our purposes, it turns out to be easier to calculate the scalar and vector potentials, rather than the electric and magnetic fields directly. The electric and magnetic fields themselves evaluated at the retarded time have a more complex time dependence [See, for example, Ref. 17, pp. 422-3] because the equations satisfied by the E&M fields have source terms that involve space-time derivatives of the source charge density and currents.

The other feature of E&M potentials we need to take into account is the so-called "gauge freedom." In short, because the electric and magnetic fields depend only on derivatives of the scalar and vector potentials, we can always add terms to the potentials to make calculations easier as long as the terms do not affect the resulting E and B fields. (This is analogous to the freedom to choose a convenient zero point for potential energy. Your results shouldn't depend on the choice of the zero point.) The bottom line is that the relationship between the source characteristics and the scalar and vector potentials can be slightly different depending on how you invoke that gauge freedom. The details will not be important here, and in any case, the electric and magnetic fields turn out to be independent of the choice of gauge.

The space and time dependence of the potentials (written generically as *A*) and their relationship to their sources (*S*) are described by wave equations. In the so-called Lorenz gauge[17, 18, 19] these wave equations look like:

$$\left(\frac{\partial^2}{\partial x^2}+\frac{\partial^2}{\partial y^2}+\frac{\partial^2}{\partial z^2}-\frac{1}{c^2}\frac{\partial^2}{\partial t^2}\right)A(x,y,z,t)=S(x,y,z,t) \ . \qquad (21)$$

As you probably know, such equations predict that E&M waves propagate (in empty space) with the universal speed $c$. Quite generally, the solutions of this kind of wave equation are given in terms of integrals of the sources evaluated at appropriate retarded times. For the case of E&M vector and scalar potentials, we have

$$\vec{A}(\vec{R},t) = \frac{\mu_0}{4\pi} \int \frac{\vec{J}(\vec{r},t_r)}{|\vec{R}-\vec{r}|} d^3r = \frac{\mu_0}{4\pi} \sum_i \frac{q_i \vec{v}_i(t_{ri})}{|\vec{R}-\vec{r}_i|} \qquad (22)$$

and

$$\Phi(\vec{R},t) = \frac{1}{4\pi\varepsilon_0} \int \frac{\rho(\vec{r},t_r)}{|\vec{R}-\vec{r}|} d^3r = \frac{1}{4\pi\varepsilon_0} \sum_i \frac{q_i}{|\vec{R}-\vec{r}_i|}, \qquad (23)$$

where the first set of source terms are integrals over a continuous current density (current per unit area) $\vec{J}$ and continuous charge density (charge per unit volume) $\rho$. The second set of source terms are the appropriate forms for a collection of point charges whose speeds are small compared to the speed of light.[20] In those equations, $\vec{R}$ is the position vector of the observation point, $\vec{r}$ is the position vector of the source point (evaluated at $t_r$), $t_r$ is the so-called retarded time $t_r = t - |\vec{R}-\vec{r}|/c$, and $c$ is the wave propagation speed (assumed here to be the usual speed of light $c$).

> **Aside**: The relationships between sources and potential functions show up in theories of many kinds of waves, including those in acoustics, plasmas, and quantum field theory. These relationships should be part of your everyday physics "toolbox."

As mentioned in the previous section, the retarded time in essence takes into account the time it takes the "message" to travel from the source point to the observation point. In other words, what one observes at the observation point at time $t$ is determined by what happened at the source point at the earlier time $t_r$.

With this whirlwind tour of E&M, we now have seen almost everything we need to develop our model of gravitational waves. Let me once again emphasize that Einstein's GR uses a rather different formulation: in GR the effects of gravity show up as curvature of space-time, which is represented mathematically by a tensor. Our analogy based on E&M provides us with a vector model of gravitational waves—the "natural" way to describe gravitation waves before GR was invented.

To calculate gravitational waves in analogy to the theory of electromagnetic waves, you simply replace the "coupling constants" (the parameters that link the source terms—masses and charges—to the potentials and fields) as follows:

$$\frac{1}{4\pi\varepsilon_0} \to -G \text{ and equivalently } \frac{\mu_0}{4\pi} \to -\frac{G}{c^2}. \tag{24}$$

$\varepsilon_0$ and $\mu_0$ are the usual permittivity and permeability of free space. In addition, charge is replaced with mass/energy and electric current is replaced with mass/energy current (mass/energy flow per unit time). The minus signs are needed because the gravitational interaction between ordinary masses is always (as far as we know) attractive. Hence, the gravitational potential energy decreases as two "like" masses approach each other.

**Exercise:** Start with Coulomb's Law for electrostatics and use the substitutions suggested in the previous paragraph. What is the result? Explain why the minus sign is needed to get the result you might expect from your knowledge of Newton's Law of Gravitation.

In analogy with E&M, let's introduce gravitational scalar and vector potentials, which, in the Lorenz gauge, satisfy the analogs of Eqs. (22) and (23) for a system of discrete masses:

$$\vec{A}_G(\vec{R},t) = -\frac{G}{c^2} \sum_i \frac{m_i \vec{v}_i}{|\vec{R}-\vec{r}_i|} \tag{25}$$

$$\Phi_G(\vec{R},t) = -G \sum_i \frac{m_i}{|\vec{R}-\vec{r}_i|}, \tag{26}$$

where $m$ is the mass and $m\vec{v}$ is the mass current. Each contribution to the right sides of Eqs. (25) and (26) is evaluated at the appropriate retarded time.

**Exercise:** For a static situation, the retarded time is not relevant (explain why) and Eq. (26) can be used to express the gravitational scalar potential without worrying about the retarded time. Go back to your introductory physics book and see if that result is consistent with what the textbook describes for the gravitational potential energy function.

In analogy with E&M waves, you might expect the oscillating gravitational radiation field, which I shall label as $\vec{g}_{rad}$ (in analogy with $\vec{g}$ for the local gravitational field in static situations), to be transverse to the wave propagation direction and to have an amplitude that falls off as $1/R$, the usual spatial dependence for the amplitude of waves far from their sources. In general, the local gravitational field $\vec{g}$ (gravitational force per unit mass) plays the role equivalent to that of the electric field of E&M.

It turns out that we can save ourselves considerable algebra by using a different gauge choice for the calculation. For the purposes of describing gravitational waves, it turns out to be expeditious to use the Coulomb gauge[17, 18] in place of the Lorenz gauge because in the Coulomb gauge the gravitational radiation field $\vec{g}_{rad}$ depends only on the components of the vector potential $\vec{A}_{G\,trans}$ that are transverse to the observation direction:

$$\vec{g}_{rad} = -\frac{\partial \vec{A}_{G\,trans}}{\partial t}. \tag{27}$$

Note that Eq. (27) does not involve the scalar potential, in contrast to Eq. (19), which involves both the scalar and vector potentials. Of course, the fields should be independent of the choice of gauge. The Coulomb gauge is convenient for our purposes because it allows us to focus just on the transverse components of the field. The essential point is that if we are interested only in the transverse components of $\vec{g}_{rad}$ we need to calculate only the transverse components of $\vec{A}_G$ (or equivalently, the transverse components of the time derivative of $\vec{A}_G$).

**Aside:** If you find the gauge arguments a bit confusing and arcane, you are in good company. It took many physicists several decades after James Clerk Maxwell published his landmark paper on electrodynamics (1865) to sort out gauge issues and settle on the mathematical techniques that are appropriate in various situations. If you are interested in understanding more about the Coulomb gauge, see Ref. 21, which discusses the details of the connections between the scalar and vector potentials in the Lorenz gauge and in the Coulomb gauge.

**Aside**: Another argument for the Coulomb gauge: the basic idea[18] behind the Coulomb gauge is that the current can be separated into a longitudinal component (along the direction of observation) and a transverse component (perpendicular to the direction of observation). It turns out that the fields produced by the longitudinal current components are proportional to the gradient of the time derivative of the scalar potential. The crucial point is that far from the source, the scalar potential contributions fall off as the inverse square of the distance $R$ between the source and observer while the radiation fields fall off only as $1/R$. Hence, at large distances from the source, the longitudinal current contributions can be ignored compared to the transverse components.

**Aside:** If you are skeptical about using the Coulomb gauge (and you should be), you could also arrive at the results by using the well-known Liénard-Wiechert potentials for point charges, the relativistic versions of Eqs. (25) and (26). (See your favorite intermediate-level E&M textbook or do a Google search.) You will need to evaluate both the scalar and the vector potentials. You should find that the "longitudinal" contribution to $\vec{g}_{rad}$ from the scalar potential cancels the longitudinal contribution from the vector potential and you are left with just transverse components.

After these formal preliminaries, we are finally ready to write down the expression for the gravitational vector potential in the Coulomb gauge:

$$\vec{A}_{G\,trans}(\vec{R},t) = -\frac{G}{c^2}\left(\sum_i \frac{m_i \vec{v}_i(t_{ri})}{|\vec{R}-\vec{r}_i|}\right)_{trans}. \tag{28}$$

The gravitational field $\vec{g}_{rad}$ calculated from Eqs. (27) and (28) has units of N/kg.

**Exercise:** Verify the claim about the units for the gravitational field.

In analogy with Eq. (20), we also expect to find a type of gravitational field akin to the magnetic field of E&M:

$$\vec{B}_G = \vec{\nabla} \times \vec{A}_G, \tag{29}$$

to which we shall return in Section V.

**Discussion Question:** What are the units for the gravitational equivalent of the magnetic field? Why don't introductory physics books talk about $\vec{B}_G$ but they spend a lot of time on E&M's $\vec{B}$?

**Wave Energy**

There is yet one more aspect of waves that we need to consider. As you learned in your physics courses, waves carry energy. Where does that energy come from? The "obvious" answer is that the energy must come from the source. The "obvious" answer might be called into question if there is some mechanism that is pumping energy into the source, for example in a laser or a radio transmitter or your cell phone. However, for the gravitational wave situation we want to describe, the binary orbiting objects can be considered as isolated from other interactions, so any energy that is carried away by the waves must result in a decrease in energy of the orbiting pair.

> **Exercise**: Cite a few examples from everyday life that indicate that waves (of any type) transport energy. For each example, what is the source of the wave energy?

To express energy carried away by the gravitational waves, we again proceed in analogy with electromagnetism. In E&M, the power per unit area in the direction of propagation can be written as

$$S_{EM} = \frac{c}{2}\left(\varepsilon_0 E^2 + \frac{1}{\mu_0}B^2\right), \qquad (30)$$

where $S_{EM}$ is the magnitude of the electromagnetic Poynting vector.[17, 18] If your E&M course did not cover E&M waves, the expression (except for the factor of $c$) should be familiar from the description of the energy density (energy per unit volume) stored in electric and magnetic fields (for example, in a capacitor or a long inductor).

> **Exercise**: Let's assume that $\frac{1}{2}(\varepsilon_0 E^2 + B^2/\mu_0)$ in Eq. (30) is appropriate for the energy density associated with **oscillating** electric and magnetic fields that constitute E&M waves. Imagine constructing a (stationary) cylinder whose axis lies along the direction of propagation of the wave and assume that the cross-section area of the cylinder is small enough that the amplitudes of the fields are constant across that area. (1) How much E&M wave energy is contained within that cylinder at some instant of time? (2) How long does it take for that energy to cross the "end cap" area of the cylinder? (3) Use those results to argue that Eq. (30) (with the factor of $c$) describes the power per unit area (perpendicular to the wave propagation direction) carried by the wave.

In practice, what one usually measures is the power averaged over several wave periods. Using $B = E/c$ for E&M plane waves, the time-averaged Poynting vector for a wave traveling in the $z$ direction can be expressed as

$$\langle \vec{S}_{EM} \rangle = c\varepsilon_0 \langle \vec{E}^2(t) \rangle \hat{z}, \qquad (31)$$

where the angular brackets indicate a time-average over several periods of oscillation. If the time dependence is sinusoidal, the averaging yields $\frac{1}{2}E^2$, where $E$ is the amplitude of the oscillating field.

> **Exercise:** Verify the statement about averaging $E^2(t) = E^2 \sin^2 \omega t$ over one (or more) period(s) of oscillation.

**Exercise:** Verify that the units of magnetic field and electric field are consistent with $B = E/c$.

We assume that the power per unit area carried by a gravitational wave is given by expressions analogous to Eqs. (30) and (31) involving the time-average of a gravitational Poynting vector. For a wave traveling in the $z$ direction, we have

$$\left\langle \vec{S}_G \right\rangle = \frac{c}{8\pi G} \left\langle \vec{g}_{rad}^2 + c^2 \vec{B}_{G\,rad}^2 \right\rangle \hat{z}. \tag{32}$$

In Section V, I shall show that $\left|\vec{g}_{rad}\right|^2 = c^2 \left|\vec{B}_{G\,rad}\right|^2$ for monochromatic gravitational radiation fields far from their sources, so I may write the time-averaged gravitational Poynting vector as

$$\left\langle \vec{S}_G \right\rangle = \frac{c}{4\pi G} \left\langle \vec{g}_{rad}^2(t) \right\rangle \hat{z}. \tag{33}$$

**Exercise**: Verify that the units in Eq. (33) work out as expected.

Although this analogy seems straightforward, there are in fact some subtleties associated with expressing gravitational field energy density both in the classical Newtonian formulation and in GR.[12, 13, 22] In particular, if we followed the substitution rules in Eq. (24), we would end up with negative energy carried by the gravitational radiation field. That minus sign had convinced Maxwell (in his famous 1865 paper on the dynamics of electric and magnetic fields) that he could not build a dynamical vector field formulation for gravity as he had done for E&M. Maxwell firmly believe that E&M energy was carried by stresses and strains in a medium (the "luminiferous ether") and he could not imagine how a medium could have a negative energy associated with it. Of course, we now know, thanks to Einstein, that the "luminiferous ether" is not necessary, so the argument about a medium has lost its force. Nevertheless, Maxwell's conclusion seems to have driven physicists away from vector theories of gravity. For now, let's ignore those issues and assume that disturbances ("wiggles") in the gravitational field will cause oscillations of masses subject to that field, thereby increasing their energy. Hence the waves must carry positive energy. That statement has its most famous formulation in Richard Feynman's "sticky bead" argument. See https://en.wikipedia.org/wiki/Sticky_bead_argument>.

**Meta-moment:** This section pushed the limit of what most undergraduate physics majors know about E&M. Were you comfortable being pushed near (or over) that limit? Why or why not? What does that tell you about how you deal with knowledge that is just beyond what you are comfortable with?

## V. VECTOR POTENTIAL FOR ORBITING BINARIES

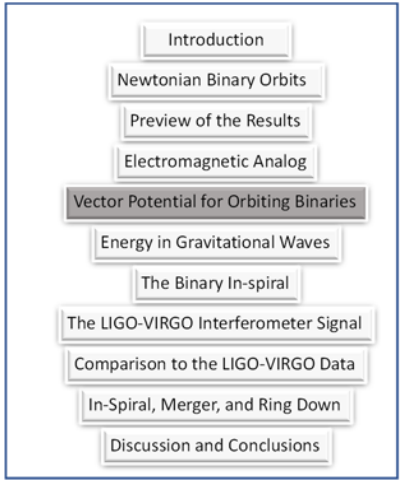

We are now ready to assemble the ingredients to find the gravitational vector potential for the specific case of orbiting binaries, the type of system responsible for the gravitational waves recently observed by the LIGO-Virgo collaboration. For a binary system of two point masses (objects whose spatial extent is very small compared to their separation), Eq. (28) becomes

$$\vec{A}_G = -\frac{Gm_a \vec{v}_a(t_{Ra})}{c^2 R_a} - \frac{Gm_b \vec{v}_b(t_{Rb})}{c^2 R_b}, \quad (34)$$

where the velocities and positions are evaluated at the appropriate retarded times.

We now take advantage of the fact that $r_a$ and $r_b$ are very small compared to $R$ for any reasonable astrophysical situation. (We shall later see how this plays out for the LIGO-Virgo observations.) This fact will allow us to replace $t_{Ra}$ and $t_{Rb}$, implicit in Eq. (34), with $t_R$ plus small correction terms. To find those terms, we first express $R_a$ in terms of $R$, the distance between the center of mass and the observation point, and $r_a$ using the Law of Cosines:

$$R_a = \sqrt{R^2 + r_a^2 - 2\vec{R}\cdot\vec{r}_a} = R\sqrt{1 + \left(\frac{r_a}{R}\right)^2 - 2\frac{\vec{R}\cdot\vec{r}_a}{R^2}}. \quad (35)$$

Invoking $R \gg r_a, r_b$, introducing the unit vector $\hat{R} = \vec{R}/R$, and employing the binomial expansion yields

$$R_a \approx R - \hat{R}\cdot\vec{r}_a \text{ and } R_b \approx R - \hat{R}\cdot\vec{r}_b, \quad (36)$$

keeping terms only through $\hat{R}\cdot\vec{r}_a$ and $\hat{R}\cdot\vec{r}_b$, which will be sufficient for the purposes of this calculation.

**Exercise:** Verify the details leading to Eq. (36).

V2  2019When we apply Eq. (34) to the case of the orbiting binaries, we need to recognize that the retarded times appear in several places in Eqs. (6)-(9). First, let's express the velocities in terms of the center of mass retarded time $t_R$ plus small correction terms. To see how this works, we will use a Taylor series expansion of $\vec{v}_a(t_{R_a})$ about the time $t_R = t - R/c$, that is, in powers of the difference

$$t - \frac{R_a}{c} - (t - \frac{R}{c}) = \frac{1}{c}(R - R_a) = \frac{1}{c}\left[\hat{R} \cdot \vec{r}_a\right] \equiv \Delta t_a, \tag{37}$$

where the penultimate equality in Eq. (37) follows from Eq. (36).

Using the standard Taylor series expansion formula yields

$$\vec{v}_a(t - R_a/c) = \vec{v}_a(t - R/c) + \left.\frac{d\vec{v}_a}{dt}\right|_{t-R/c} \Delta t_a + \ldots. \tag{38}$$

Eqs. (6) and (7) give

$$\frac{d\vec{v}_a}{dt} = -\omega^2 \vec{r}_a \tag{39}$$

with an analogous expression for the derivative of $\vec{v}_b$.

Assembling these results and using $R \approx R_a \approx R_b$ in the denominator of Eq. (34) (all retarded time "corrections" in the denominator lead to terms that go as $1/R^2$, which you may ignore for radiation fields), you should find that the gravitational vector potential can be written as

$$\vec{A}_G(\vec{R},t) = -\frac{G}{c^2 R}\left[m_a \vec{v}_a(t_R) + m_b \vec{v}_b(t_R)\right] + \frac{G\omega^2}{c^3 R}\left[m_a \vec{r}_a(\hat{R} \cdot \vec{r}_a) + m_b \vec{r}_b(\hat{R} \cdot \vec{r}_b)\right]. \tag{40}$$

In Eq. (40), all of the terms are evaluated at the same retarded time $t_R = t - R/c$.

**Exercise:** Check the details leading to Eq. (40), which is a critical result for our analysis.

The first term on the right side of Eq. (40) can be ignored because the net linear momentum of the orbiting binaries is zero (or more generally a constant, so when we take the time derivative to get $\vec{g}_{rad}$, it makes no contribution). The second term goes as $1/R$, hence we will have $g_{rad} \sim 1/R$, characteristic of a radiation field. By way of contrast, we note that for the E&M vector potential produced by two oppositely charged objects (an electric dipole), the equivalent first term does not vanish.

The second term involves products of the masses and the square of the distances from the center of mass, that is, they are related to the "quadrupole moment tensor" of the mass distribution $Q_{ij} = \sum_k m_k \left(3 r_i r_j - r^2 \delta_{ij}\right)$ for masses $k = 1, N$ with $i, j = x, y, z$ and $r^2 = \sum_i r_i^2$. This result indicates that gravitational radiation is primarily quadrupole (rather than dipole) radiation. (This result also follows from the GR treatment of gravitational waves.) In more advanced courses in E&M, you will learn about a systematic approach, the so-called multi-pole expansions of the scalar and vector potentials, to these kinds of calculations.

Eq. (40) can be put into a more perspicacious form by using Eq. (1) to replace $\vec{r}_a$ and $\vec{r}_b$ in favor of $\vec{r}$:

$$\vec{A}_G(\vec{R},t) = \frac{G\omega^2}{c^3 R} \frac{m_a m_b}{m_a + m_b} \left[\vec{r}(\hat{R}\cdot\vec{r}_a) - \vec{r}(\hat{R}\cdot\vec{r}_b)\right] = \frac{G\omega^2}{c^3 R} \eta M \left[\vec{r}(\hat{R}\cdot\vec{r})\right]. \tag{41}$$

As a reminder, I note that the mass ratio $\eta$ is defined to be

$$\eta \equiv \frac{m_a m_b}{\left(m_a + m_b\right)^2} \tag{42}$$

and the total mass is $M = m_a + m_b$. As I mentioned in Section III, the combination $\eta M$ is the reduced mass of the system.

**Exercise**: Check the algebra leading from Eq. (40) to Eq. (41).

We now use Eq. (10) to write the vector potential as

$$\begin{aligned}\vec{A}_G(\vec{R},t) &= \frac{G\eta M \omega^2 r^2}{c^3 R}\left[\left(\hat{x}\cos\omega t_R + \hat{y}\sin\omega t_R\right)\left(\hat{R}\cdot\hat{x}\cos\omega t_R\right)\right] \\ &= \frac{G\eta M \omega^2 r^2}{c^3 R}\left[\hat{x}\, n_x \cos^2\omega t_R + \hat{y}\, n_x \cos\omega t_R \sin\omega t_R\right] \\ &= \frac{G\eta M \omega^2 r^2}{2c^3 R}\left[\hat{x}\, n_x (1 + \cos 2\omega t_R) + \hat{y}\, n_x \sin 2\omega t_R\right],\end{aligned} \tag{43}$$

where $n_x = \hat{R}\cdot\hat{x}$. Note that the choice of the orientation of the x-y axes has eliminated $\hat{R}\cdot\hat{y}$ terms. (See Fig. 1.) The time-dependent terms indicate that the gravitational vector potential and hence the field oscillate with a frequency which is **twice the orbital frequency**, a characteristic of quadrupole radiation from an orbiting binary system: after half of an orbital period, the second moment returns to its initial value (since both *x* and *y* coordinates are reversed), indicating that the second moment oscillates with twice the orbital frequency.

To find the transverse components of the vector potential along the direction of observation, we express it in spherical coordinates ($R, \theta, \phi$) using the standard expressions

$$\hat{x} = \sin\theta_i \cos\phi \, \hat{R} + \cos\theta_i \cos\phi \, \hat{e}_1 - \sin\phi \, \hat{e}_2 = \sin\theta_i \, \hat{R} + \cos\theta_i \, \hat{e}_1 \tag{44}$$

$$\hat{y} = \sin\theta_i \sin\phi \, \hat{R} + \cos\theta_i \sin\phi \, \hat{e}_1 + \cos\phi \, \hat{e}_2 = \hat{e}_2 \tag{45}$$

$$\hat{z} = \hat{R}\cos\theta_i - \hat{e}_1 \sin\theta_i \tag{46}$$

$$n_x = \sin\theta_i \cos\phi = \sin\theta_i . \tag{47}$$

The second equality in each of Eqs. (44)-(47) follows from the choice of $\phi = 0$ in the orientation of the *x-y* coordinate system relative to the observation direction. As mentioned previously, we will focus on just the $\hat{e}_1 = \hat{\theta}$ and $\hat{e}_2 = \hat{\phi}$ terms—the transverse terms.

Using the appropriate trigonometric relations to re-express the terms in Eq. (43) yields

$$\vec{A}_{G\,\text{trans}} = \frac{G\eta M \omega^2 r^2}{2c^3 R} \sin\theta_i \left[\cos\theta_i (1+\cos 2\omega t_R)\hat{e}_1 + \sin 2\omega t_R \, \hat{e}_2\right]. \tag{48}$$

We now take the time derivative of the transverse vector potential in Eq. (48) to get the gravitational radiation field

$$\boxed{\vec{g}_{\text{rad}} = -\frac{\partial \vec{A}_{G\,\text{trans}}}{\partial t} = \frac{G\eta M \omega^3 r^2}{2c^3 R}[\sin 2\theta_i \sin 2\omega t_R \, \hat{e}_1 - 2\sin\theta_i \cos 2\omega t_R \, \hat{e}_2].} \tag{49}$$

This is our fundamental result for the gravitational radiation field for the orbiting binaries.

**Exercise:** Rev up your trigonometry engine and check the details leading from Eq. (41) to Eqs. (48) and (49).

**Aside:** Let us pause to make a general comment about the calculation: it is important to note that the use of "differential" retarded times in the analysis leads to expressions that depend on the structure of the source and the relative velocities of the source components. If we had evaluated Eq. (34) with only the retarded time associated with the center of mass $t_R = t - R/c$ and had ignored the differences in retarded times for the two masses, we would not have found any gravitational waves. The same comments apply to the importance of differential retarded time for E&M waves.[17] In more general treatments of waves and their sources, the distribution of "charges" and "currents" (either electrical or mass) is treated in what is called a multi-pole expansion. For a systematic treatment of waves, the multi-pole expansion is essential. What we worked through in the previous pages is essentially a "dummies" version (in the spirit of the *XXXX for Dummies* book series) of the first stage of a multi-pole expansion. See Ref. 17, pp. 454-458; and Ref. 18, pp. 391-401 and Chapter 16 for details of the multi-pole expansions.

**Exercise**: Our hypothetical pre-1915 physicist could use Eq. (49) to estimate the amplitude of the gravitational radiation field from a particular source. Obviously, to get the largest wave amplitude we want to have large, nearly equal masses (to keep $\eta$ as large as possible) orbiting at a high frequency and not too far away from Earth. The only obvious candidates in 1915 would be binary stars in the Milky Way. To see how the numbers work out, let's assume that the sum of the masses is 20 solar masses and that the orbital separation between the stars is about the sum of their radii (estimated by scaling from the Sun's radius). Let's also assume that the binary stars are about 1000 light years away, well within the Milky Way. Show that with those numbers the orbital frequency $\omega$ is about $3 \times 10^{-4}$ rad/sec and the amplitude of $g_{rad}$ is about $10^{-16}$ m/s$^2$. Such a small acceleration at such a low frequency would have discouraged almost anyone from thinking about detecting gravitational waves.

Next, we tackle the details of the gravimagnetic field $\vec{B}_G$ associated with the gravitational radiation. That field is found from Eq. (48) by using the spherical coordinate form of the curl:

$$\nabla \times \vec{A} = \frac{1}{r \sin \theta} \left[ \frac{\partial}{\partial \theta} (\sin \theta A_\phi) - \frac{\partial A_\theta}{\partial \phi} \right] \hat{r} + \frac{1}{r} \left[ \frac{1}{\sin \theta} \frac{\partial A_r}{\partial \phi} - \frac{\partial}{\partial r} (r A_\phi) \right] \hat{\theta} \\ + \frac{1}{r} \left[ \frac{\partial}{\partial r} (r A_\theta) - \frac{\partial A_r}{\partial \theta} \right] \hat{\phi}. \tag{50}$$

Recalling that there is $R$ dependence in the retarded time $t_R = t - R/c$, we see that derivatives with respect to $R$ can be replaced by time derivatives using the standard chain rule:

$$\frac{\partial f(t - R/c)}{\partial R} = -\frac{1}{c} \frac{\partial f}{\partial t}. \tag{51}$$

**Exercise:** The result linking a spatial derivative along the observation direction to a time derivative of the oscillating field is a very general relationship for waves. Construct an explanation of why this works that would convince an introductory physics student that it applies to waves with sinusoidal temporal and spatial oscillations.

Using Eq. (51) and dropping terms that go as $1/R^2$, we find

$$\vec{B}_{G\,rad} = \vec{\nabla} \times \vec{A}_G = \frac{1}{c} \left[ \left( -g_{rad\,\phi} \right) \hat{e}_1 + \left( g_{rad\,\theta} \right) \hat{e}_2 \right]. \tag{52}$$

The important result is that $\vec{g}_{rad}^2 = c^2 \vec{B}_{G\,rad}^2$ just as $E^2 = c^2 B^2$ for E&M waves.

**Meta-moment**: Although this section was relatively straightforward, you needed to pay attention to lots of detail and it may not have been obvious what the next step would be. Did that make you feel nervous or anxious about your ability to carry out this kind of calculation? I admit that it took three or four iterations of this calculation to get it into a coherent form. This is fairly typical of scientific work: you often need several iterations of a calculation or an experiment until you find the method that works most smoothly. Have you had that kind of experience in any of your STEM courses?

## VI.   ENERGY CARRIED BY GRAVITATIONAL WAVES

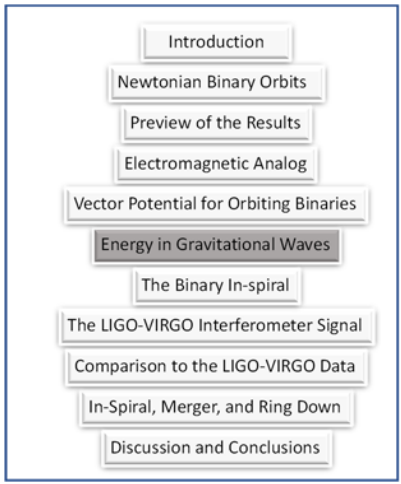

We now want to find the power per unit area carried away by the gravitational wave as expressed by the time-average of the gravitational radiation Poynting vector. As we mentioned in the section on the orbital dynamics of binaries, as the system loses energy to gravitational waves, the binaries spiral in towards each other, increasing their orbital angular frequency, which, as we shall see, leads to enhanced gravitational wave energy loss. Using Eqs. (32) and (49) and straightforward algebra and trigonometry, we obtain

$$\langle S_G \rangle = \frac{c}{4\pi G} \langle \vec{g}_{rad} \cdot \vec{g}_{rad} \rangle$$
$$= \frac{G\omega^6 (\eta M)^2 r^4}{32\pi c^5 R^2} \left[ \sin^2 2\theta + 4\sin^2 \theta \right]. \tag{53}$$

I have made use of $\langle \cos^2 2\omega t \rangle = \langle \sin^2 2\omega t \rangle = 1/2$, where the average is taken over one (or more) oscillation periods of the wave. I shall return to a discussion of the angular distribution in the final section of the tutorial and simply note here that it differs from the angular distribution derived from the linear form of GR.

**Exercise:**  Check the details leading to Eq. (53).

To find the total gravitational power radiated, we need to integrate Eq. (53) over the surface of a sphere of radius *R*, centered on the source. In spherical coordinates the differential area on the surface of a sphere is given, as usual, by $R^2 \sin\theta \, d\theta \, d\phi$, so we need to integrate

$$R^2 \int_0^{2\pi} d\phi \int_0^{\pi} (\sin^2 2\theta + 4\sin^2 \theta)\sin\theta \, d\theta = \frac{64\pi}{5} R^2. \tag{54}$$

**Exercise:** Check the integration results given in Eq. (54).

Assembling the results, we find that the total gravitation wave energy radiated per unit time can be written as

$$\frac{dE_G}{dt} = \frac{c}{32\pi G} \frac{G^2(\eta M)^2 r^4 \omega^6}{c^6} \frac{64\pi}{5} = \frac{2}{5} \frac{G(\eta M)^2 r^4 \omega^6}{c^5}. \tag{55}$$

Eq. (55) is the crucial result needed to account for the physics of the decay of the binary orbits. Except for the overall numerical factor, this result is in agreement with the result calculated from the linear version of GR,[8, 12, 13] including the dependence on $G$, $M$, $r$, $\eta$ and $\omega$. The GR numerical factor for the last term in Eq. (55) is 32/5. (Note that I could get the GR and E&M-like models to agree if I increased the E&M GW <u>amplitude</u> by a factor of 4. I will return to that issue in Section XI.) The difference is attributable to the difference in angular distributions of the radiated power between the E&M-like model and the linear GR results. Please note that the $R^2$ factors have cancelled out due to the "conspiracy" between the $R^2$ dependence of the surface area of a sphere (a geometric effect) and the $1/R^2$ dependence of the energy per unit area (due to how the interactions depend on $R$).

This set of exercises is not essential to the main argument, but it may help convince you that Eq. (55) has the correct form.
**Exercise:** The results stated in Eq. (55) are surprisingly simple and demand a simple explanation. Let's repackage Eq. (55) to provide such an explanation. Ignoring numerical factors, we may write

$$\frac{dE_G}{dt} \approx \left(\eta M \, r^2 \omega^2 \omega\right)^2 \frac{G}{c^5} \tag{56}$$

The term in parentheses on the right in Eq. (56) is the square of the rate at which kinetic energy, proportional to $\eta M \, r^2 \omega^2$, is carried around the orbit. (Recall that $\eta M$ is the reduced mass of the system and $\omega = 2\pi/T$, where $T$ is the orbital period.) The remaining term ($G/c^5$) can be viewed as the reciprocal of a fundamental unit of gravitational radiation luminosity $L_0$ (energy per unit time) with the numerical value $L_0 = c^5/G = 3.628 \times 10^{52}$ J/s in SI units. So, Eq. (55) can be viewed[12] as the rate at which kinetic energy is carried around the orbit multiplied by the ratio of that rate to the fundamental luminosity unit $L_0$:

$$\frac{dE_G}{dt} \approx (\text{KE transfer rate}) \times \frac{(\text{KE transfer rate})}{L_0}. \tag{57}$$

To appreciate the size of the fundamental unit of gravitational radiation luminosity, we note that the Sun's luminosity is about $10^{26}$ J/s and the electromagnetic luminosity of the Milky Way galaxy is about $10^{36}$ J/s.

**Exercise**: Provide arguments to justify that $\eta M\, r^2 \omega^2 \omega$ is proportional to the rate of kinetic energy transport around the orbit by the binary objects.

**Exercise:** Show that $c^5/G$ has the units of luminosity (energy per unit time). Verify the stated numerical value of $L_0$ in standard SI units.

**Exercise**: There is yet another formulation that provides further insight about the physical content of Eq. (55). We can think of $\eta M\, r^2$ as the quadrupole moment of the binary system. Since the quadrupole moment tensor components oscillate at $2\omega$, $\eta M\, r^2 \omega^3$ is proportional to the third time derivative of the quadrupole moment tensor, that is to $\dddot{Q}_{ij}$. Hence, we may write

$$\frac{dE_G}{dt} \approx \sum_{i,j} \left|\dddot{Q}_{ij}\right|^2, \tag{58}$$

a result which is the crucial ingredient in the linear GR theory of gravitational waves.

**Exercise**: Come up with a better explanation of the physics contained in Eq. (55) and share it with the author of this tutorial. He is always looking for better ways of understanding important results.

**Meta-moment:** Do you routinely go through the exercise of trying to understand a complicated mathematical expression in physical terms? Have you had the experience of using physical arguments to guess at what a mathematical representation should look like, even without doing a detailed derivation?

## VII. THE BINARY IN-SPIRAL

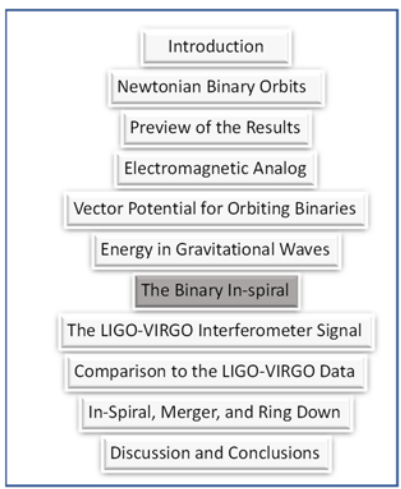

As mentioned previously, the binary masses spiral in towards each other as gravitational waves carry energy away from the system. As a consequence, the orbital frequency increases. The increased orbital frequency leads to an increase in both the frequency and amplitude of the gravitational wave radiation field as described by Eq. (49). To see how the rate of change of the frequency $\dot{\omega}$ is related to the source properties, we construct an expression that indicates how the orbital frequency changes as the binary system radiates energy. Let's start with the time derivative of Kepler's Third Law (Eq. (4)) to obtain

$$\dot{\omega} = -\frac{3}{2}\frac{\sqrt{GM}}{r^{5/2}}\dot{r}. \tag{59}$$

The time rate of change of $r$, the separation between the orbiting masses, can be related to the rate of energy loss, Eq. (55), (rewritten here with the numerical factor indicated by $N$)

$$\frac{dE_G}{dt} = N\frac{G(\eta M)^2 r^4 \omega^6}{c^5}. \tag{60}$$

We set the rate of change of the orbital energy, Eq. (11), equal to the negative of Eq. (60) to find

$$\dot{U}_{total} = \frac{Gm_a m_b}{2r^2}\dot{r} = -N\frac{G(\eta M)^2 r^4 \omega^6}{c^5}. \tag{61}$$

Again, using Kepler's Third Law to replace $\omega$ in favor of $r$, we find

$$\dot{r} = -\frac{2NG^3 M m_a m_b}{c^5 r^3} = \frac{-2N(GM)^3 \eta}{c^5 r^3}, \tag{62}$$

which when combined with Eq. (59) gives

$$\dot{\omega} = \frac{3N\eta(GM)^{5/3}\omega^{11/3}}{c^5}. \tag{63}$$

Solving for $(\eta N)^{3/5} M$ yields

$$(\eta N)^{3/5} M = \frac{c^3}{3^{3/5} G}\left(\dot{\omega}\omega^{-11/3}\right)^{3/5}. \tag{64}$$

Note that Eq. (64) involves the combination $\eta^{3/5} M$, which is called the chirp mass. The name is appropriate because an oscillatory signal with increasing frequency and amplitude is called a "chirp." Eq. (64) tells us that if we determine $\dot{\omega}(t)$ and $\omega(t)$ from the GW observations, we can calculate the chirp mass of the binary system.

Next let's rewrite Eq. (62) in a form that will be useful in what follows. First we recall that the distance scale for any relativistic theory of gravity is set by the so-called Schwarzschild radius.[23] We also remember that the Schwarzschild radius can be defined as the distance from a mass $M$ at which the escape speed is equal to the speed of light:

$$r_S = \frac{2GM}{c^2}. \tag{65}$$

Using Eq. (65) in Eq. (62), we find the previously stated Eq.(16),

$$\dot{r} = \frac{-\eta N c}{4}\left(\frac{r_S}{r}\right)^3, \tag{66}$$

a remarkably simple expression. In essence, the Schwarzschild radius sets the scale for the in-spiral. As $r \to r_S$, the rate of change of the separation increases dramatically. Note that for a given mass separation $r$, the E&M-like model ($N = 2/5$) gives a smaller rate of change of $r$ than does the linear GR model ($N = 32/5$).

We can calculate the orbits during the in-spiral using the following procedure. First recall that Eq. (66) may be integrated directly to find $r(t)$, given previously in Eq. (17):

$$r^4(t) = r_i^4 - N\eta r_S^3 c(t - t_i), \tag{67}$$

where $r_i$ is the initial mass separation at time $t_i$. If $r_i$ is a few times the Schwarzschild radius for the total mass, the in-spiral will be easily noticeable. It is helpful to rewrite Eq. (67) so that the separation is expressed in units of the Schwarzschild radius:

$$\frac{r(t)}{r_S} = \left[\left(\frac{r_i}{r_S}\right)^4 - \frac{N\eta c(t - t_i)}{r_S}\right]^{1/4}. \tag{68}$$

Note that the factor inside the square brackets in Eq. (68) can become negative for sufficiently large $t$, so there is an obvious limit to the application of this model.

**Exercise**: Use Eq. (68) to find $t_*$ for which $r(t_*) = r_S$.

Note that one could, alternatively, integrate Eq. (63) to find $\omega(t)$ and start the calculation based on an initial frequency rather than an initial mass separation. Once we have $r(t)$ and $\omega(t)$ (from Kepler's Third Law) we can easily find $v_{a,b} = \omega r_{a,b}$ to get the speeds of the masses relative to the center of mass.

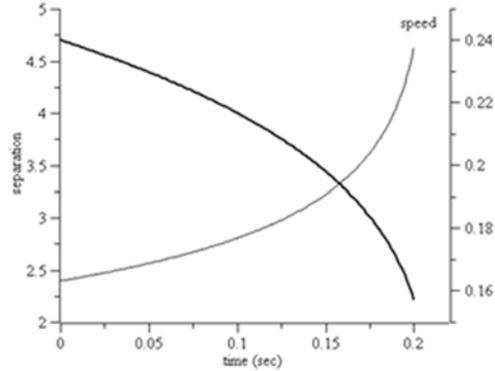

Fig. 3. The mass separation (thick curve, left axis, in units of the Schwarzschild radius $r_S$) and the speed (thin curve, right axis, in units of $c$) of one of the masses relative to the center of mass of the system plotted as a function of time for the conditions stated in the text, with $N = 32/5$.

The results of such a calculation are shown in Fig. 3 for $N = 32/5$ (the linear GR value) with $m_a = m_b = 35 M_\odot$, where $M_\odot$ is the mass of the Sun. The initial separation is $4.7 r_S$. We see that strong orbit in-spiral occurs over a period of only a few tenths of a second and the speed $v$ of the masses relative to the center of mass is a few tenths of the speed of light. From $r(t)$ (or $\omega(t)$), we can calculate the gravitational radiation field from Eq. (49). A video simulation of the in-spiraling binaries is available at the URL given in Ref. 24.

**Exercise**: Use a spreadsheet or simple computer program to reproduce the results shown in Fig. 3. Try different values of the parameter $\eta$. What is the physical explanation for what happens when $\eta$ changes?

**Meta-moment**: This section showed you the calculations needed to derive the results described in section III. Was the wait worth the effort? Do you prefer to have derivations first or to have the results given first followed by the derivation? Why?

## VIII. THE LIGO-VIRGO INTERFEROMETER SIGNAL

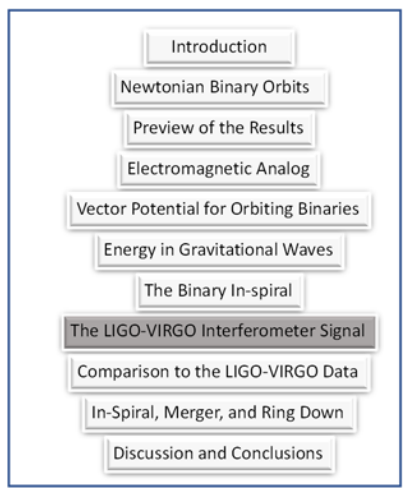

Our next task is to relate the oscillating gravitational field $\vec{g}_{rad}$ to what LIGO-Virgo observes. Then in the following section, we will discuss how the observed signal can be used to find information about the presumed gravitational wave source: orbiting binaries.

The LIGO-Virgo observatory uses a relatively complex Michelson optical interferometer[25] with perpendicular arms of length $L$ and Fabry-Perot cavities in each arm, to measure the gravitational wave "strain" $\Delta L / L$, where $\Delta L$ is the change in arm length induced by a gravitational wave. We will develop an argument that relates $\vec{g}_{rad}$ to $\Delta L / L$. To keep the argument as simple as possible, we will focus attention on just one arm of the interferometer, which is sufficient for connecting $\vec{g}_{rad}$ to $\Delta L / L$.

**Exercise**: Review the operation of a Michelson interferometer as described in your favorite optics book (or on the web). In particular, pay attention to what happens to the interferometer's interference pattern when one of the mirrors moves relative to the beam splitter.

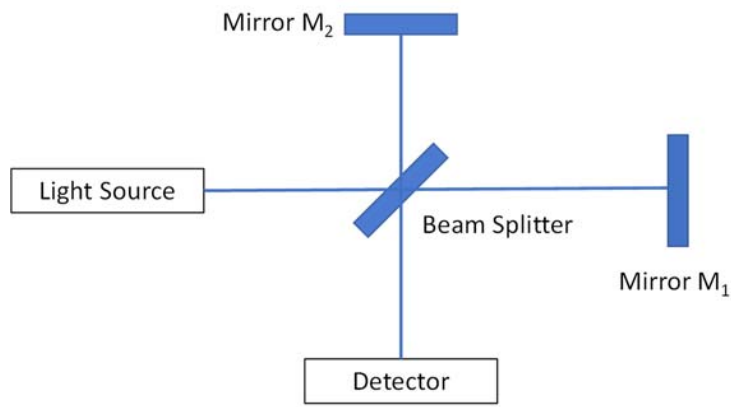

Fig. 4. A typical Michelson interferometer. Light from the source (usually a laser) hits a beam splitter. Part of the light is reflected from the beam splitter and is then reflected by mirror $M_2$. The other part of the beam goes through the beam splitter and reflects from mirror $M_1$. After hitting the beam splitter again, parts of the two beams combine and hit the Detector, where an interference pattern is observed. If one of the mirrors moves along an "arm" of the interferometer, the interference pattern will shift on the Detector.

**Exercise:** Take a quick look at Ref. 25 to see how the LIGO interferometer is constructed.

Interferometric gravitational wave detectors like LIGO-Virgo respond to differential movements of parts of the apparatus (in particular, the interferometer mirrors and beam splitter in an interferometer arm). Differential movement occurs only if the gravitational wave interacts with the mirror and the beam splitter at (slightly) different times. (For a nice explanation of how the interferometer responds to gravitational waves, see Ref. 8, pp. 289-293.) In addition, the details relating the gravitational wave function $\vec{g}_{rad}$ to the strain are complicated by the geometry of the wave propagation direction and polarization directions (given by the $\hat{e}_1$ and $\hat{e}_2$ terms in Eq. (49)) relative to the interferometer arms—the so-called antenna pattern.[26, 27] The simple model used here will minimize those complications. There are two essential steps to the calculation: (a) Project the gravitational radiation field $\vec{g}_{rad}$ along the line joining the mirror and the beam splitter. (b) Take into account the time it takes the wave to travel from one object to the other.

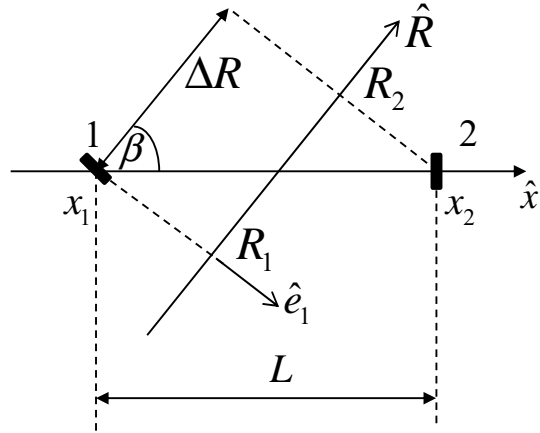

Fig. 5. The simple model for relative displacements of two masses, 1 (the beam splitter) and 2 (the mirror), in one arm of length $L$ in a LIGO-Virgo interferometer. The gravitational wave is traveling in the $\hat{R}$ direction. The non-orthogonal $(x, R)$ coordinates are given by $(x_i, R_i)$ with $i = 1, 2$. $\hat{e}_1, \hat{x},$ and $\hat{R}$ lie in the plane of the figure. Assume that the masses are free to move only in the $x$ direction (along the interferometer arm). $\hat{e}_1 \cdot \hat{x} = \sin \beta$, $\Delta R = R_2 - R_1$ and the length of the arm is given by $L = x_2 - x_1$. $\hat{e}_2$ points into the plane of the page.

In the simple model, the plane gravitational wave is propagating along the $\hat{R}$ direction and we assume that the mirror and beam splitter are free to move only along $\hat{x}$. (See Fig. 5.) Let's assume that the polarization unit vector $\hat{e}_1$ is in the $x$-$R$ plane and that $\hat{e}_2 \cdot \hat{x} = 0$. In that case, Eq. (49) expresses the acceleration of mass 1 (the beam splitter) as

$$\ddot{x}_1 = \vec{g}_{rad} \cdot \hat{x} = \frac{G\eta M \omega^3 r^2}{2c^3 R} \sin\beta \sin 2\theta_i \sin(2\omega t), \tag{69}$$

with a similar expression for mass 2 (the mirror).

**Exercise**: Check the dependence on $\beta$ and $\theta_i$ given in Eq. (69). Verify that the units in Eq. (69) are units of acceleration.

Both the mirrors and the beam splitter in the LIGO-Virgo interferometer are mounted as pendula[25] and are free to oscillate along the interferometer arm direction. The mirror and beam splitter can be modeled as linear, driven, damped oscillators. Since the gravitational wave frequency (about 30-200 Hz for the LIGO-Virgo observations) is much larger than the natural oscillation frequencies of the beam splitter and mirror and since the pendulum damping is negligible during the time of interaction with the short bursts of gravitational waves observed so far, we find that the displacements of the beam splitter and the mirror in the $x$ direction are given approximately by

$$x(t) \approx -\frac{\vec{g}_{rad} \cdot \hat{x}}{(2\omega)^2}. \tag{70}$$

The minus sign in Eq. (70) simply tells us that the displacement and forcing term are $\pi$ radians out of phase for the sinusoidal driving of an object at a frequency far above its natural oscillation frequency. The 2 in the denominator reminds us that the wave frequency is twice the binary orbital frequency.

**Exercise**: Hold the end of the string of a string-and-mass pendulum in your hand. Observe the natural oscillation frequency and period of the pendulum. Then move your hand back-and-forth horizontally with a period long compared to the oscillation period of the pendulum. Notice that your hand and the pendulum bob move in phase. Now move your hand back-and-forth very rapidly (compared to the natural oscillation frequency). You should notice that your hand and the pendulum bob move 180° out of phase and the faster you move your hand, the smaller the pendulum oscillation amplitude. Are those results consistent with Eq. (70)?

**Exercise:** Review the physics of a sinusoidally-driven damped oscillator and justify the result in Eq. (70) for the case when the driving frequency (here $2\omega$) is much larger than the natural oscillation frequency of the oscillator.

Using Eqs. (69) and (70), we may write the time-dependent relative displacements of the two masses as

$$\Delta x_2(t) - \Delta x_1(t) = -\frac{G\eta M \omega r^2}{8c^3 R} \sin\beta \sin 2\theta_i \left[\sin(2\omega t) - \sin(2\omega(t - \Delta t))\right], \quad (71)$$

where I introduced the wave travel time from $R_1$ to $R_2$ as $\Delta t = \Delta R/c$. Note that we may drop the minus sign by shifting the overall (unimportant) phase of the gravitational wave by $\pi$. The time-dependent terms in Eq. (71) become

$$\left[\sin(2\omega t) - \sin(2\omega(t - \Delta t))\right] \approx 2\omega \Delta t \cos(2\omega t), \quad (72)$$

where I used standard trigonometric expressions and invoked the small angle approximation: $\sin(2\omega \Delta t) \approx 2\omega \Delta t$. We can use $2\omega \Delta t \ll 1$ because the gravitational wave travel time from $x_1$ to $x_2$ is small compared to the wave period, which is certainly the case with the LIGO-Virgo observatory whose interferometer arms are 4 km long.

**Exercise**: Work through the details expressed in Eq. (72). Show explicitly that $2\omega \Delta t \ll 1$ for the LIGO-Virgo observatories.

The differential $x$ displacement $\Delta L$ of the two masses is then given by

$$\Delta L = \Delta x_2(t) - \Delta x_1(t) = \frac{G\eta M \omega^2 r^2}{4c^3 R} \Delta t \sin\beta \sin 2\theta_i \cos(2\omega t). \quad (73)$$

**Exercise**: Check the details leading to Eq. (73).

Let's now use Kepler's Third Law to replace $r$ in favor of $\omega$ in the previous expression:

$$\Delta L = \frac{\eta(GM)^{5/3} \omega^{2/3}}{4c^4 R} L \sin\beta \cos\beta \sin 2\theta_i \cos(2\omega t), \quad (74)$$

where also I used $\Delta t = \Delta R/c = L\cos\beta/c$.

Eq. (74) provides the important information: The relative displacement oscillates at twice the orbital frequency of the source and the amplitude depends on the combination $\eta^{3/5} M$, the chirp mass, multiplied by $\omega^{2/3}$, exactly the dependence predicted by linear GR.

It is straightforward to show that the differential displacement for the orthogonal arm of the interferometer (in the *y* direction) is the same as Eq. (74) with the opposite overall sign. In other words, the gravitational wave creates antisymmetric displacements in the two arms and that leads to a shift in the interference pattern observed by the interferometer's detector. The detector signal is proportional to the relative strain given by $(\Delta L_x - \Delta L_y)/L$. (For the sake of simplicity, we assume that the two arm lengths are the same.)

We find that the interferometer strain signal $h(t)$ is given by:

$$h(t) = \frac{\Delta L_x(t) - \Delta L_y(t)}{L} = \frac{\eta (GM)^{5/3} \omega^{2/3}}{2c^4 R} \sin\beta \cos\beta \sin 2\theta_i \cos(2\omega t). \tag{75}$$

This form of the strain signal has the same dependence on $G$, $M$, $\omega, \eta, c,$ and $R$ as the expression derived from GR.[13, 28] As noted previously, the angular dependence is different. Appendix B provides a derivation of $h(t)$ for an arbitrary GW propagation direction.

If we use the Schwarzschild radius $r_S = 2GM/c^2$, the amplitude of the strain signal in Eq. (75) can be rewritten in a surprisingly simple form:

$$|h(t)| = \frac{\eta}{8} \frac{r_S}{r} \frac{r_S}{R} \times \text{angular factors} \tag{76}$$

**Exercise**: Verify that Eq. (76) follows from Eq. (75).

As we have noted previously, the Schwarzschild radius $r_S$ sets the distance scale for gravitational waves. The $r_S/R$ factor shows how the amplitude falls off with distance from the source to the observation point, and the $r_S/r$ term tells us that the amplitude is largest when the mass separation approaches the Schwarzschild radius. If we can make reasonable estimates of the angular factors, we can use Eq. (76) and the measured amplitude of the strain signal ($\sim 10^{-21}$ for the first LIGO-Virgo observation) to estimate the distance of the binaries from Earth. (See Section IX.)

**Calculating the Gravitational Waveform**

Next let's see how we can calculate the gravitational waveform given an initial choice for the masses and their separation, aiming for parameter values that yield signals displaying the in-spiral events observed by LIGO-Virgo. In Section IX, we will describe how to estimate the masses from the observed LIGO-Virgo waveform.

The Schwarzschild radius, Eq. (15), is used to set the scale for the following calculational steps:

1. Select an initial value of the separation of the masses that is a few times the combined Schwarzschild radii of the binary objects. (The strong in-spiral of the binary objects begins when the separation of the masses is a few times the combined Schwarzschild radii.)

2. Use Eq. (68) to find $r(t)$ for a sequence of time values.

3. From $r(t)$, use Kepler's Third Law to find $\omega(t)$.

4. Use that value of $\omega$ in Eq. (75) and repeat steps 3 and 4 for the next time value. Please note that this method implicitly assumes that $\omega$ does not change significantly during an orbital cycle. This assumption is violated as the binary objects get close to each other. See below for a more general method.

5. Plot the results.

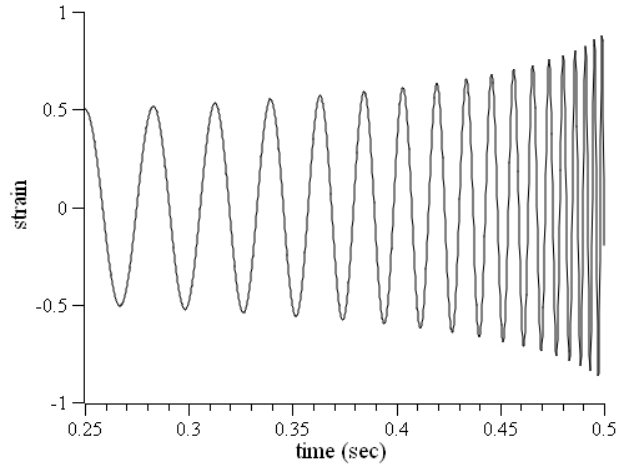

Fig. 6. The gravitational wave signal (arbitrary units) calculated as described in the text and plotted as a function of time with total mass $M = 340$, $N = 2/5$, and $\eta = 0.25$ for the E&M-like model, which is equivalent to $M = 65$ and $\eta = 0.247$ (close to the LIGO-Virgo GW150914 results) with $N = 32/5$ for the linear GR theory. Masses are multiples of the solar mass. The initial value of the mass separation is $1.7\, r_S$ (1760 km) for the E&M-like model and $5.3\, r_S$ (1000 km) for the linear GR theory. The final value of the mass separation is $1.2\, r_S$ (1200 km) for the E&M-like model and $3.6\, r_S$ (680 km) for linear GR.

Results of this calculation are shown in Fig. 6, which plots the strain wave signal as a function of time for conditions close to those for the LIGO-Virgo GW150914 observations, described in detail in the following section. [Note that "GW150914" means "Gravitational Wave (event) on September 14, 2015.] The signal consists of an oscillatory waveform with gradually increasing frequency and amplitude as the masses undergo an in-spiral due to the loss of orbital energy to the emitted gravitational waves. Note that both the E&M-like calculation and the linear GR theory give **exactly** the same waveform if the same product $(\eta N)^{3/5} M$ is used in the two theories.

As mentioned above, the assumption of slow variation of the orbital (and wave) frequencies is violated as the binary objects get close to each other. The correct way to handle a chirp signal (or any signal in which the frequency changes with time) is to replace $\cos(2\omega(t)t)$ with $\cos(2\Phi(t))$, where $\Phi(t)$ is the integrated phase of the orbital motion:

$$\Phi(t) = \int_{t_o}^{t} \omega(\tau) d\tau . \tag{77}$$

Let's see how that works out for a GW in-spiral signal. Eq. (63) provides us with a differential equation, which is straightforward to solve for $\omega(t)$ (See the Appendix A for details.):

$$\omega(t) = \frac{1}{\sqrt{2}\, \tau_{\text{S-chirp}}} \left( N \frac{t_c - t}{\tau_{\text{S-chirp}}} \right)^{-3/8} , \tag{78}$$

where $t_c$ is the "coalescence time," or more precisely, the time at which the orbital frequency becomes infinite in our model. $\tau_{\text{S-chirp}} = 2GM_{\text{chirp}}/c^3$ is the "Schwarzschild chirp time," the time for a signal with speed $c$ to travel a distance of one Schwarzschild radius associated with the chirp mass of the system.

.

In Appendix A, I show how to use Eq. (78) to express the integrated orbital phase:

$$\Phi(t) = -\frac{1}{5}\left(\frac{16}{N^{3/5}}\right)^{5/8} \left[\frac{t_c - t}{\tau_{\text{S-chirp}}}\right]^{5/8} + \Phi_c . \tag{79}$$

$\Phi_c$ is the phase when the two objects coalesce and can be adjusted to fit the data. Overall the integrated orbital phase is dimensionless.

**Exercise:** Show how to derive Eq. (78) from Eq. (63) and then show that Eq. (79) follows. See the Appendix A for some assistance.

**Exercise:** Show that the numerical factor in Eq. (79) is $(2/5)^{5/8}$ for $N = 32/5$, the general relativity value of $N$.

The integrated phase method is particularly important for smaller mass binary objects such as neutron stars. With the smaller masses, the radiated GW power is smaller than for larger masses and the GW waveform is within the LIGO-Virgo observatory frequency range for a longer time. For example, for a pair of neutron stars, each with $m \approx 1.4 M_\odot$, the observed GW signal lasts for tens of seconds. With those longer signals, the "error" due to using $\omega(t)$ in place of the more appropriate $\Phi(t)$ becomes quite noticeable. In 2017, the LIGO-Virgo collaboration detected signals associated with the binary orbit merger of two neutron stars.[29]

**Meta-moment:** This section required developing a model of the LIGO-Virgo interferometer and how it responds to a gravitational wave. This is often viewed as an "advanced" topic even for experts in GR. You should take a moment to reflect about your ability to arrive at this result with just a standard background in undergraduate physics, with some guidance, of course.

## IX. COMPARISON TO THE LIGO-VIRGO DATA

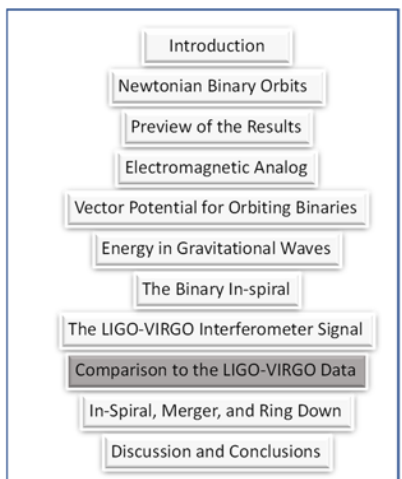

Now we will compare the results of our calculations with the LIGO-Virgo results. The comparison of the E&M-like calculations (and those from the linear form of GR) with the LIGO-Virgo results is complicated by two factors. First, the observatories are sensitive only to gravitational waves with frequencies between approximately[1,25] 20 and 1000 Hz and contain moderately complex frequency filtering to reduce noise in the signal. Second, both the E&M-like calculation of gravitational waves and the linear GR theory do not depend on and do not contain any information about the binary objects when they "collide." So, our analysis must be limited to comparing signals before the two objects merge and seeing what can be learned from the frequency of the waves and the frequency's evolution with time. (However, see Section X for a phenomenological description of the merger.) In what follows we will compare the results of the E&M-like/linear GR calculation with the numerical GR results as given by LIGO-Virgo.

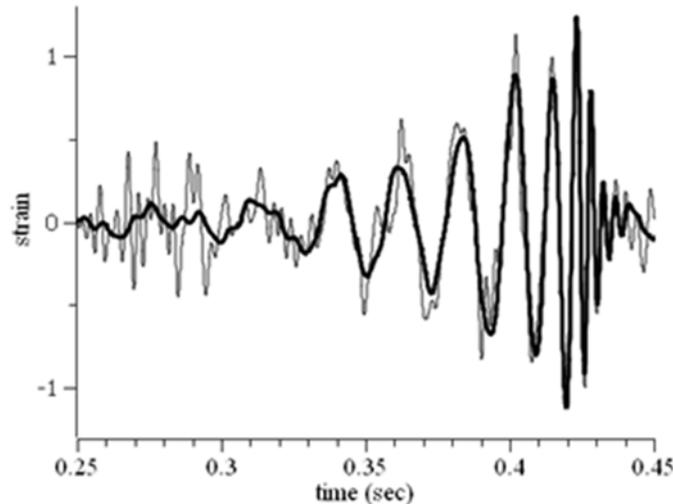

Fig. 7. The gravitational wave strain data[30] (multiplied by $10^{21}$) from the LIGO event GW150914 plotted as a function of time (in sec). The thin curve is the raw data from the LIGO interferometer in Hanford, WA. The thick curve is the waveform from a full numerical GR calculation taking into account the frequency response of the detector system. See Ref. 1.

The LIGO-Virgo website[30] for the GW150914 event provides both the raw data for the observed waveforms and a numerical GR gravitational strain waveform that takes into account the frequency response of the instrument. Both are shown in Fig. 7. The signal consists of an oscillatory part with gradually increasing frequency and amplitude as the binary masses approach each other. Then a sudden increase in the wave frequency and a rapid decay in amplitude at $t \approx 0.42$ sec indicate the merger and coalescence of the two objects.

**Exercise:** Visit the LIGO Open Science[30] website and find the data plotted in Fig. 7. Produce your own graph of the data.

Let's see how we might extract information from the waveform.[5, 31,32, 33] This is a critical argument in understanding the LIGO-Virgo collaboration's conclusion that the detected gravitational waves came from in-spiraling black holes. First, we note that Eq. (64) indicates that the chirp mass $\eta^{3/5}M$ can be extracted from the data by finding $\omega(t)$ and $\dot{\omega}(t)$. From the LIGO-Virgo data for GW150914 and using the linear GR value $N = 32/5$, we find that the chirp mass is about $30 M_\odot$. Assuming as a first approximation that the two masses are equal ($\eta = 1/4$), we find that the total mass is about $70 M_\odot$. The detailed numerical GR analysis, which includes information gleaned from the merger and ring-down, yields 36 and $29 M_\odot$ (with uncertainties of about $5 M_\odot$) for the two masses in a frame of reference at rest with respect to the source. The GW frequencies observed on Earth are red-shifted due to cosmological expansion of the universe. That redshift is characterized by a parameter z. For the GW150914 event, the source red-shift is given by $z \approx 0.09$. That means that the mass determined from the signal detected on Earth is related to the mass in the source-frame by $M_{detector} = (1+z) M_{source}$, yielding detector-frame masses of 39 and 31 $M_\odot$. Since we are comparing theoretical predictions with the detector signal, we should use the detector-frame values.

As mentioned previously, the waveforms from the two theories agree exactly as long as the product $(\eta N)^{3/5} M$ is the same. For the E&M-like model ($N = 2/5$), we find that the total mass $M \approx 340 M_\odot$. However, different values of $M$ will lead to different results for the mass separation $r(t)$ for a given orbital frequency.

> **Exercise:** Use the GlowScript program BinaryInSpiral to see a simulation of the in-spiral and a plot of the corresponding signal that would be observed by LIGO-Virgo. The URL for the program is http://www.glowscript.org/#/user/rhilborn/folder/Public/program/BinaryInSpiral.
>
> If you run your cursor over the gravitational waveform data generated by the GlowScrript program, you can read off the coordinates of points on the curve. Use those readings to carry out the analysis described in the preceding paragraph. (To check the results of your analysis, look at the program code to see what mass values were used in the simulation.)

Once we have an initial estimate for the masses, we may calculate the gravitational wave strain signal numerically using the method described in Section VIII. The results presented here were calculated using *Mathematica*, but even a spreadsheet is adequate for this calculation. To match the LIGO-Virgo numerical GR results, we first note that the initial wave frequency of the strain signal in Fig. 7 is about 30 Hz, corresponding to a wavelength of about $10^7$ m, much larger than the LIGO-Virgo interferometer arm length. Recalling that the orbital frequency is half the wave frequency, we choose the initial separation of the two orbiting masses to give an orbital frequency $\omega/2\pi \approx 15$ Hz. The initial separation of the two masses is about $10^6$ m for the LIGO-Virgo masses and about $1.7\times10^6$ m for the E&M-like model mass values. Those separations are about a factor of 10 smaller than the emitted wavelength. The zero of time is adjusted to match the LIGO-Virgo data.

**Exercise**: Verify the numerical values stated in the previous paragraph. Compare the separation distances with the size of a typical star like the Sun.

The final separation is chosen so that the calculation matches the time in the LIGO-Virgo signal at which "merger and ring down" begins (at about 0.42 sec in Fig. 7). That time corresponds approximately to the time at which the largest amplitude of the detected waveform occurs.[34] For the purposes of this paper, we plot the E&M-like model/linear GR waveform with an amplitude that is a multiple of its amplitude at $t = t_0$, where $t_0$ is the initial time point, and we scale that amplitude to match the LIGO-Virgo initial numerical GR amplitude. Eq. (75) tells us that the amplitude increases as the 2/3 power of the frequency.

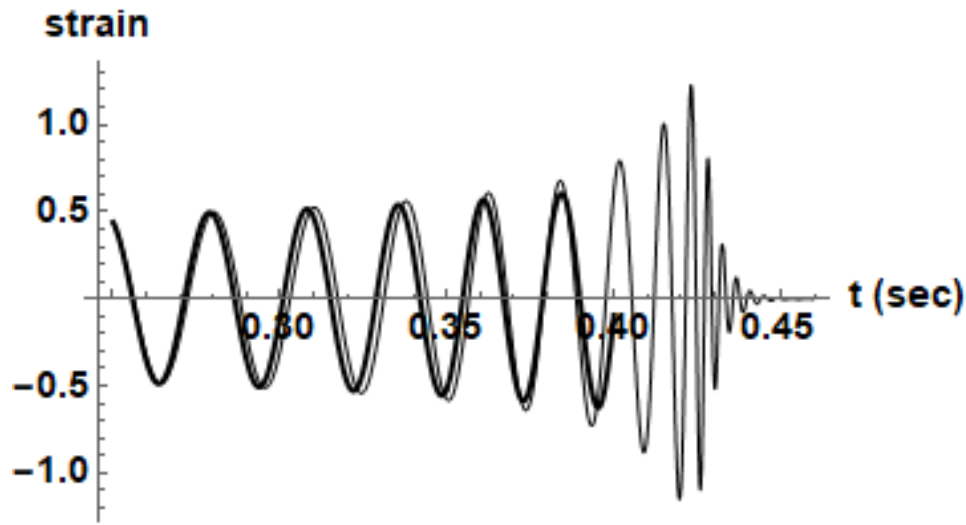

Fig. 8. The "unfiltered" numerical GR calculation[30] of the LIGO Hanford GW150914 strain signal (thin curve) (multiplied by $10^{21}$) plotted as a function of time for the GW150914 event as the signal would have been seen by a detector without noise or frequency response limitations, along with the predictions of the E&M-like model/linear GR model (thick curve) with parameters described in the text. The initial time was taken to match the LIGO plot and the theoretical calculation is cut off at $t = 0.4$ s where the coalescence and ring-down begin, indicated by a rapid increase in the wave frequency.

Figure 8 displays the results of the E&M-like model/linear GR theory along with the "unfiltered" numerical GR results, removing the frequency response effects of the apparatus. The agreement is surprisingly good given that the E&M-like model/linear GR theory includes none of the GR nonlinearities or relativistic dynamics effects. Those results match the numerical GR results except in the last few milliseconds, at which time the merger and ring-down occur. For the E&M-like model, the mass separation at about 0.42 sec is $r \approx 1.37 r_S \approx 1.4 \times 10^6$ m and $v/c \approx 0.3$ for each of the two masses. For linear GR, $r \approx 4.1 r_S \approx 7.8 \times 10^5$ m and $v/c \approx 0.16$.

How do we conclude that the orbiting objects are black holes? Could the objects be ordinary stars? Let's start with the Sun. The radius of the Sun is about $7 \times 10^5$ km. More massive stars are even larger. Since our results show that the objects have masses about 30 times the solar mass and that they approach each other within about $10^6$ m, we can rule out ordinary stars. What about more compact astrophysical objects? White dwarf stars are thought to be limited to masses at most 1.4 times the mass of the Sun. So, they won't work. How about neutron stars, which are the collapsed cores of large stars? The radius of a neutron star is about 10 km, so that would work, but masses of neutron stars are thought to lie between 1.1 and 3 solar masses. So they won't work either. We are led to the conclusion that the orbiting objects are most likely black holes, since, of all known astronomical objects with masses on the order of 30-100 $M_\odot$, only black holes can approach within $10^6$ m.[5, 34] Of course, we must allow for the possibility of some exotic object. But by using detailed numerical GR calculations that are beyond the scope of this tutorial, the LIGO-Virgo collaboration[34] simulated a binary black hole merger and found that the details of the signal amplitude during the last moments of the in-spiral and of the coalescence are consistent with the objects' being black holes.

The data can also be used to estimate the distance from Earth to the binaries, if we make reasonable assumptions about the detector antenna pattern, the wave polarization and the binary orbit angle of inclination relative to the observing direction. The LIGO-Virgo strain signal has an amplitude of about $10^{-21}$ for the GW150914 event from which the full numerical GR analysis[34] yields a distance of about 410 Mpc with an uncertainty of about 170 Mpc [1 Megaparsec (Mpc) is about 3.3 million light-years]. The uncertainty is largely due to the unknown orbit inclination angle.

There are two ways we can estimate the source-Earth distance without using numerical GR. First, we could use Eq. (76), which relates the observed strain signal (for a given binary mass separation) to the distance $R$. The mass separation can be estimated as indicated previously from the observed wave frequency. For example, if we use the LIGO masses for the binaries and $f_{wave} \approx 200\,\text{Hz}$, we have $r \approx 1.5\,r_S$. With $|h(t)| \approx 10^{-21}$ for the GW150914 observation, we get $R \leq 130\,\text{Mpc}$. The upper limit holds if the angular factors are about equal to 1.

The second method is to estimate the total gravitation wave power (luminosity) of the binaries using Eq. (55). Then, linear GR links[13] the luminosity to the observed strain via

$$L = \frac{c^3}{16\pi G}\int |\dot{h}(t)|^2 dS = \frac{c^3}{4G} R^2 \omega_{wave}^2 |h|^2, \tag{80}$$

where $dS$ is the differential surface element and the last equality makes use of $|\dot{h}| = \omega_{wave}|h|$ for a sinusoidally periodic wave. Eq. (55) (with the linear GR numerical factor 32/5) yields $L \approx 10^{49}\,\text{J/s}$ for the LIGO-Virgo masses when $\omega_{wave} = 2\pi \times 200\,\text{rad/s}$. With $h \approx 10^{-21}$, Eq. (80) yields $R \leq 300\,\text{Mpc}$. Given the approximations used, these estimates are in generally good agreement.

**Exercise**: Compare those estimates of distances with sizes of galaxies. Were those events close by (astronomically speaking) or far away?

**Discussion Question:** The gravitational wave peak luminosity of about $10^{49}$ J/s is large (though short-lived). As mentioned previously, the Sun's (electromagnetic) luminosity is about $10^{26}$ J/s. and the luminosity of the entire Milky Way Galaxy is about $10^{36}$ J/s. What is going on with merging black holes that allows their luminosity to be so large?

**Exercise**: The LIGO-Virgo collaborative has two active interferometer sites in the United States: one in Livingston, LA and the other in Hanford, WA. The comparison between the Hanford and Livingston signals allows estimates of the spatial location of the source in the sky. The GW150914 signal arrived at Livingston about $6.9 \pm 0.5$ ms before it arrived in Hanford,[34] in rough agreement with gravitational waves traveling at speed $c$ over the approximately $D = 3000$ km "line of sight" (for gravitational waves) between Hanford and Livingston. Denoting the angle between the wave propagation direction and the line running from Livingston to Hanford as $\phi$, show that the time delay between the signals at the two detectors is given by $\Delta t = (D/c)\cos\phi$. For the GW150914 event, show that $\phi \approx 45°$.

**Exercise**: What is the longest delay time between observations in Hanford and those in Livingston that would be consistent with gravitational waves traveling at speed $c$?

> **Exercise:** Use Eq. (49) and the values for the GW150914 binaries to find the value of $R$ for which the magnitude of $g_{rad}$ is about 10 m/s². Is that a safe distance at which to observe the two orbiting black holes?

## X.   UNDERSTANDING THE IN-SPIRAL, MERGER, AND RING DOWN

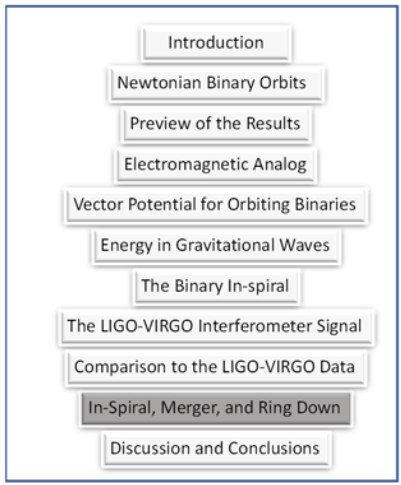

What determines when the final in-spiral and merger events occur? As mentioned previously, neither the E&M-like model developed in this tutorial nor the linear GR theory contain information about the nature of the binary objects and their "merger." So, this section deviates substantially from our previous calculations. It will be more speculative, and in the terms used by scientists, more phenomenological: what kinds of plausibility arguments can we provide that give us at least a qualitative understanding of the merger and ring-down of the colliding objects?

The calculations in the previous sections indicate that the rapid collapse of the orbit occurs when the fractional change in orbital frequency per orbital period becomes significant (say, about 0.1). Once that point is reached, the system goes through 10-12 orbits in a few tenths of a second (in the blink of an eye) and then the merger is complete and the gravitational radiation stops. What is the physics behind that rapid change and merger?

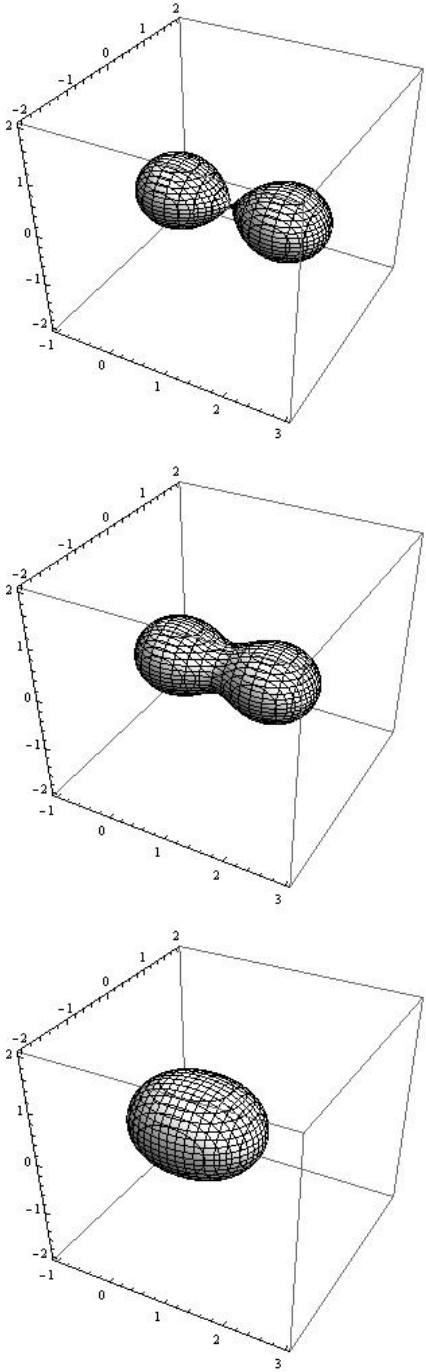

Fig. 9. Plot of the non-relativistic no-escape surface ($v_{esc}/c = 1$) for binaries with equal masses. The separation between the masses is given as a multiple of the sum of the Schwarzschild radii. $r = n r_s$ with $n = 2.0, 1.8,$ and $0.8$ from top to bottom. The masses are located at $(x, y, z) = (0, 0, 0)$ and $(n, 0, 0)$. All distances are multiples of the sum of the Schwarzschild radii. The surfaces are viewed in a frame of reference rotating at the orbital frequency.

By drawing a contour plot of the condition that the non-relativistic escape speed equals the speed of light, we gain some insight about why these conditions mark the onset of the rapid in-spiral. On the surfaces shown in Fig. 9, we have

$$1 = v_{\rm esc}/c = \sqrt{\frac{2Gm_a}{c^2 r_a} + \frac{2Gm_b}{c^2 r_b}} = \sqrt{\frac{r_{Sa}}{r_a} + \frac{r_{Sb}}{r_b}}, \tag{81}$$

where $r_{a,b}$ are the distances from masses $m_{a,b}$ to the surface point and $r_{Sa,b}$ are the associated Schwarzschild radii. It is important to note that the contour plots were made for a static situation using just the standard Newtonian gravitational potential without taking into account relativistic effects or the nonlinearities inherent in GR. We must also remember that the black holes continue to orbit one another as their separation decreases. Nevertheless, the surfaces provide some insight into what happens in the final merger.

I point out that in general relativity there are several possibilities for an "effective surface" for a single black hole.[13] The radii of those surfaces depend on the mass and spin of the black hole. None of them take into account the presence of the other black hole, the relativistic corrections due to the high speeds of the orbiting objects, or the nonlinear effects due to the strong spacetime curvature. These comments highlight how difficult it is to describe the merger of two black holes quantitatively and that we need to view our crude picture with a good dose of skepticism.

Figure 9 shows contour plots of the no-escape surface for the mass separations $r = 2r_S, 1.8r_S,$ and $0.8r_S$ for the case of equal mass binaries. (There is nothing special about those separations. They were chosen to give a rough idea of the evolution of the system as the "merger" takes place.) With $r = 1.8 r_S$ for example, both masses clearly lie within the common no-escape surface. An external observer "sees" a single, but non-spherical object. The system continues emitting gravitational waves until the no-escape surface becomes almost spherical as shown qualitatively in Fig. 9 with $r = 0.8 r_S$. Although this model does not give us a dynamical quantitative treatment or firm quantitative predictions for ring down,[35] it does make it plausible that the gravitational radiation, which requires a quadrupole structure, should cease when the two objects merge.

We can think of the last picture in Fig. 9 as a distorted ("perturbed") black hole and then ask how it relaxes to a spherical shape as it emits GWs along the way. Examination of Fig. 7 indicates that the ring-down, which begins at about $t = 0.42$ sec, shows damped oscillations with a frequency considerably higher than the frequencies associated with the binary orbit. The higher frequency suggests that a new mode of oscillation begins as the two objects merge. In fact, GR predicts that a "perturbed" black hole will oscillate in what are called quasi-normal modes.[36] (Quasi-normal means that the frequencies are represented by complex numbers, whose real part gives the oscillation frequency and whose imaginary part gives the damping of the oscillations.)

It is plausible[5] that the period of the ring-down oscillations is roughly the time it takes a signal to travel across the orbit whose mass separation, we have argued, is a few times the combined Schwarzschild radii of the two objects. Consequently, we expect the ring-down angular frequency $\omega_{ring}$ to satisfy $\omega_{ring} \tau \approx 1$, where $\tau = n r_S / c$. For $n = 3$, that estimate[37] gives a ring-down wave frequency (twice the GW ring-down oscillation frequency) about equal to 170 Hz for LIGO-Virgo GW150914 mass values and 320 Hz for the E&M-like model masses. These results are reasonably close to the observed 200-250 Hz ring-down wave frequency.[1] (See Fig. 7.) Of course, this agreement is most likely fortuitous since the estimate does not include the strong gravitational effects in the region around the black holes. Strong gravity (strong space-time curvature in GR) certainly influences the characteristics and propagation of the gravitational waves.

Since the observed ring-down wave oscillations are rapidly damped, a reasonable model for the strain during the ring-down phase is

$$h(t) = A e^{-(t-t_s)/\delta t} \cos(2\omega_{ring} t + \phi), \tag{82}$$

where the ring-down start time $t_s$, the damping parameter $\delta t$, and the phase $\phi$ can be adjusted to mimic the observed wave form. Results are shown in Fig. 10.

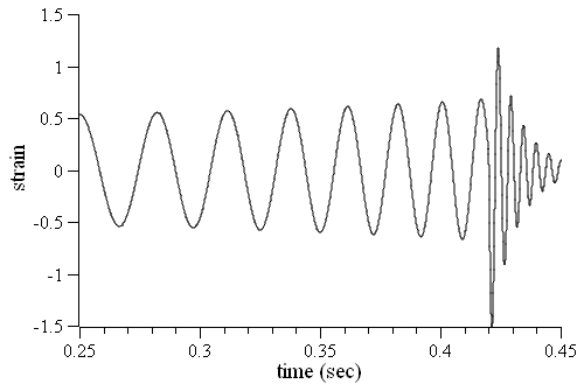

Fig. 10.  The E&M-like model/linear GR strain signal, given by Eq. (75), with a phenomenological model, given by Eq. (82) for the ring-down phase, which begins at about $t = 0.42$ sec.  The ring-down start time $t_s = 0.42$ sec is estimated from the LIGO-Virgo signal shown in Fig. 7. The amplitude and the phase $\phi$ of the ring-down part of the signal are chosen to match the amplitude and phase of the pre-merger signal at the ring-down start time.  The damping parameter, $\delta t = 0.01$ sec, is chosen to approximate the LIGO-Virgo observations and the numerical GR results.

## XI.  DISCUSSION AND CONCLUSIONS

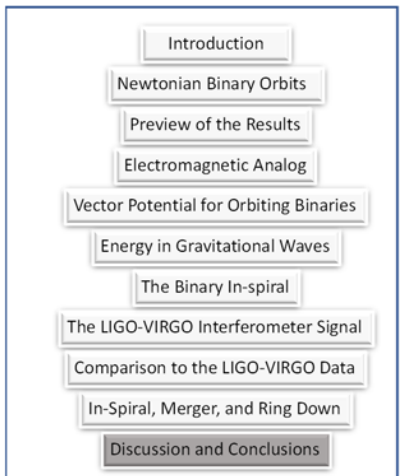

Using analogies with electromagnetism, we have seen that a relatively simple model of the gravitational radiation emitted by orbiting binaries reproduces many of the features of the linear form of GR calculations for the same system.  The simple model does surprisingly well in describing the detected gravitational waveform observed by LIGO-Virgo and the numerical GR calculations of those waveforms except for their final moments.  The simple model is independent of the nature of the orbiting masses, so it can say nothing about what happens when the objects collide.

However, the polarization properties of the gravitational waves, the spatial distribution of the emitted energy, and the overall numerical factor for the luminosity in the E&M-like model are notably different from the linear GR results.  Although the physics underlying the two approaches is essentially the same (viz. oscillating mass quadrupole moments produce gravitational waves). Tthe E&M-like model treats $\vec{g}(t)$ as the dynamical field, while in GR the metric tensor is the dynamical field. Those differences lead to differences in the angular distributions of emitted energy. In particular for the circular binary orbit situation, Eq. (53) tells us that in the E&M-like model there is no gravitational radiation in the direction perpendicular to the plane of the binary orbit, while the results from linear GR are[13, 28]

$$h_+(t) = \frac{1}{R}\frac{G^2 2m_a m_b}{rc^4}(1+\cos^2\theta_i)\cos 2\omega t_R$$
$$h_\times(t) = \frac{1}{R}\frac{G^2 4m_a m_b}{rc^4}\cos\theta_i \sin 2\omega t_R \ . \tag{83}$$

In Eq. (83), $h_+(t)$ and $h_\times(t)$ deviations of the metric tensor from a flat space-time model and directly give the LIGO-Virgo signals with oscillations patterns ("plus" and "cross" polarizations) that are spatially 45° apart with a $\pi/2$ temporal phase difference between them. The other symbols are the same as those used previously in this tutorial. As we have seen, the E&M-like model (Eq. (49)) gives two orthogonal polarizations in the $\hat{e}_1$ and $\hat{e}_2$ directions with a $\pi/2$ temporal phase shift between them.

Eq. (83) indicates that in the linear GR theory the gravitational wave power is a maximum perpendicular to the plane of the orbit. The vector theory gives zero power in that direction. There is, however, almost no information available about the orientation of the binary orbits relative to the line-of-sight for the recent LIGO-Virgo observations except for the binary neutron star event GW180817 and even in that case the interpretations of the data are ambiguous.

**Exercise:** Show that the amplitude of the strain signal in Eq. (83), ignoring the angular factors, is the same as that in Eq. (75).

**Exercise**: Compare the predictions of the E&M-like model with those of the linear GR theory for radiation emitted in the plane of the orbits $(\theta = \pi/2)$.

The differences in the polarization properties of the gravitational waves between GR and the E&M-like model, as well as in other alternative theories of gravitation,[38] can in principle be tested when additional gravitational wave observatories are active on Earth. Those other observatories will be oriented differently relative to the gravitational wave propagation direction so that comparing data among the observatories should yield information about the GW polarization. The recent detection of a gravitational wave event GW170814 at three interferometer sites (LIGO Hanford, WA, LIGO Livingston, LA, and Virgo Pisa, Italy) in principle allows a determination of the polarization properties of the waves since the interferometer arms of the Virgo detector are oriented in space differently from those of the Hanford and Livingston, which have almost identical orientations. The analysis carried out by the LIGO-Virgo collaboration focused on the possibility of scalar, vector, and tensor polarizations within the framework of metric theories of gravitational waves. In a metric theory, it is the space-time metric tensor that "oscillates" in a gravitational wave. (In our E&M-like model, it is the gravitational field that oscillates in a gravitational wave.) The LIGO-Virgo collaboration claims[39] that the GR tensor polarizations are compatible with the observations, but a metric theory with pure vector or pure scalar forms are not. However, the details of that analysis have not yet been published. As noted previously, the E&M-like model polarizations are different from those of the GR tensor theory, but it remains to be seen if the data are sufficient to distinguish between them given the relatively large uncertainties in the source sky location.

You have also seen that neither the E&M-like model or the linear GR model can account for what happens during the merger and ring-down parts of the observed signal. Using some simple physics of the no-escape surfaces for the orbiting binaries gives a plausible account, as shown in Section X, of the merger of black holes and a phenomenological explanation for the ring-down but does not provide a quantitative account of that part of the signal.

**Aside:** For the sake of completeness, I point out another set of observations that indicate the existence of gravitational waves. The detailed timing measurements of the radio-frequency pulses emitted by the Taylor-Hulse binary pulsar[40] over several decades of observations indicate that the orbital period of the pulsar is decreasing at a rate completely consistent with the orbital energy loss due to gravitational waves predicted by linear GR [Eq. (55) with $N = 32/5$]. This result seems to rule in favor of the linear GR approach. We note that the pulsar orbital period decay rate is $10^{11}$ times smaller than the decay rate observed in the black hole in-spiral observed by LIGO-Virgo. The E&M-like model analysis of the pulsar orbital period decay rate will be discussed in a separate publication.

**Aside**: As I mentioned previously, if the GW amplitudes predicted by the E&M-like theory were increased by a factor of 4, then the energy radiated by orbiting binaries would be exactly the same in both the E&M-like model and the linear GR model. (The polarization properties and angular distribution of the GWs would still be different.) One might justify a factor of 4 in Eq. (25) because we have no prior experimental data about gravitational fields produced by moving

> masses. However, it would be better if I could find some theoretical justification for that factor. Since I have not (yet) found such a justification, I will stick with the arguments that do not have such a factor. Any theoretical suggestions will be most welcome.

The important "take home message" from this tutorial is that GR, although it is our most fully tested relativistic theory of gravitation, is not needed to give reasonable accounts of many of the properties of the existing observations of gravitational waves. This comment should not be taken as downplaying the revolutionary way of thinking about gravitation that is embodied in GR. As mentioned previously, accounting for the details of the source behavior as the binary objects coalesce does require GR (in the case of two black holes) or other detailed astrophysics for the case of colliding neutron stars.

Many writers (see, for example, Ref. 41 ) claim that only GR predicts gravitational waves and that the LIGO-Virgo observations and the corresponding numerical GR calculations "prove" that GR is correct. As mentioned in the Introduction, almost any relativistic theory of gravitational interactions will predict the existence of gravitational waves. This tutorial provides a concrete realization of that statement. I have noted that GR was not the first theory to predict gravitational waves nor is it required, as you have seen, to explain many of the experimentally observed properties of gravitational waves. In fact, I have been surprised that the E&M-like model is able to account for the mass and frequency dependence of the in-spiral part of the gravitational wave signal. Nevertheless, I must emphasize that the current gravitational wave observations, except for questions about GW polarization, seem to be in good agreement with GR. There is some evidence that the GWs from the recently observed binary neutron stars favors vector polarization over tensor polarization.[42]

A catalog of the GW signals seen by the LIGO-Virgo collaboration from September 2015 to August 2017 can be found at arXiv:1811/.12907v2 "GWTC-1: A Gravitational-Wave Transient Catalog of Compact Binary Mergers Observed by LIGO and Virgo during the First and Second Observing Runs."

*Mathematica* notebooks, an Excel spreadsheet, and a GlowScript program for carrying out the numerical calculations are available upon request from the author.

> **Exercise**: Look up the data for the gravitational wave observation GW170104. Use the information in that paper to calculate the waveform, using the results developed in this tutorial.

V2 2019

**Exercise:** LIGO-Virgo has seen evidence of gravitational waves from two orbiting neutron stars.[29] Look up information on neutron stars and use the methods given in this tutorial to calculate the gravitational wave signal expected from orbiting binary neutron stars. Given the frequency response of the LIGO-Virgo interferometers, would those signals be seen with the current setup? Suppose the smallest strain amplitude that LIGO-Virgo can detect is about $10^{-22}$, what is the maximum distance for the binary neutron stars from Earth that would lead to a detectable gravitational wave signal?

**Meta-moment:** If you have worked through the details of this tutorial's arguments, you have achieved a rather amazing scientific result. You have seen in detail the basic physics behind one of the most important recent observations in physics: direct detection of gravitational waves. You have seen how to use basic physics to make quantitative predictions about the gravitational waves and you have analyzed the evidence that allows us to conclude that the detected waves were emitted by binary black holes. The observations give evidence for black holes with masses of a few tens of solar masses, ones for which evidence had been previously lacking. You should feel a great deal of satisfaction that you can understand the basic physics of these events.

## ACKNOWLEDGEMENTS

Thomas A. Moore, Joseph C. Amato, Hongbin Kim, and James Battat provided invaluable and much appreciated comments on both the content and style of draft versions of the paper on which this tutorial is based. Senaea Rose provided detailed comments and suggestions on the semi-final version of the tutorial. Barbaro Qunitero-Leyva, Niloofar Sharei, and Luca Piccotti gave useful comments on v1 of the tutorial. This research has made use of data, software and/or web tools obtained from the Gravitational Wave Open Science Center (https://www.gw-openscience.org), a service of LIGO Laboratory, the LIGO Scientific Collaboration and the Virgo Collaboration. LIGO is funded by the U.S. National Science Foundation. Virgo is funded by the French Centre National de Recherche Scientifique (CNRS), the Italian Istituto Nazionale della Fisica Nucleare (INFN) and the Dutch Nikhef, with contributions by Polish and Hungarian institutes.

a) Email address: rhilborn@aapt.org

# Appendix A
## Derivation of the Integrated Orbital Phase Signal

Equation (63), repeated here as Eq. (A1), is a differential equation, which we can solve for the orbital frequency $\omega(t)$:

$$\dot{\omega} = \frac{3N\eta(GM)^{5/3}\omega^{11/3}}{c^5} \tag{A1}$$

We solve this equation using standard calculus techniques:

$$\int_{\omega(t)}^{\infty} \frac{d\omega}{\omega^{11/3}} = \frac{3N\eta(GM)^{5/3}}{c^5} \int_{t}^{t_c} d\tau, \tag{A2}$$

where we have chosen $t_c$ (the "coalescence time") to be the time at which the orbital frequency becomes infinite (in our model). The left-side of Eq. (A2) becomes

$$\int_{\omega(t)}^{\infty} \frac{d\omega}{\omega^{11/3}} = -\frac{3}{8}\left[\frac{1}{\infty} - \frac{1}{\omega^{8/3}(t)}\right] = \frac{3}{8}\omega^{-8/3}(t). \tag{A3}$$

The right-side of Eq. (A2) is given by

$$b\int_{t}^{t_c} d\tau = b(t_c - t), \tag{A4}$$

where $b = 3N\eta(GM)^{5/3}/c^5$. It is illuminating to write the following expressions in terms of the Schwarzschild radius associated with the chirp mass ($M_{\text{chirp}} = \eta^{3/5}M$):

$$r_{\text{S-chirp}} = \frac{2GM_{\text{chirp}}}{c^2} \tag{A5}$$

and a Schwarzschild chirp time,

$$\tau_{\text{S-chirp}} = r_{\text{S-chirp}}/c, \tag{A6}$$

the time it takes a signal traveling at speed $c$ to cross a distance equal to one Schwarzschild chirp radius.

Assembling the pieces, we find

$$\omega(t) = \left(\frac{3}{8b}\right)^{3/8}(t_c - t)^{-3/8}. \tag{A7}$$

Replacing $b$ with the associated parameters yields, after a bit of algebra,

$$\omega(t) = \frac{1}{\sqrt{2}}\left(\frac{1}{N}\right)^{3/8} \frac{1}{\tau_{\text{S-chirp}}} \left[\frac{t_c - t}{\tau_{\text{S-chirp}}}\right]^{-3/8}. \tag{A8}$$

For linear GR with $N = 32/5$, we obtain

$$\omega(t) = \left(\frac{5}{2}\right)^{3/8} \frac{1}{4\tau_{\text{S-chirp}}} \left[\frac{t_c - t}{\tau_{\text{S-chirp}}}\right]^{-3/8}. \tag{A9}$$

Note that $\tau_{\text{S-chirp}}$ sets the scale for the orbital frequency and its time evolution. In these equations, we restrict ourselves to $t < t_c$.

To find the integrated orbital phase required to describe the chirp waveform, we need to integrate $\omega(t)$:

$$\begin{aligned}\Phi(t) &= \int_{t_0}^{t} \omega(\tau) d\tau = \left(\frac{3}{8b}\right)^{3/8} \int_{t_0}^{t} (t_c - t)^{-3/8} d\tau \\ &= -\frac{8}{5}\left(\frac{3}{8b}\right)^{3/8} \left[(t_c - t)^{5/8} - (t_c - t_0)^{5/8}\right].\end{aligned} \tag{A10}$$

Let's now rearrange the numerical factors and parameters into a perspicacious form:

$$\frac{8}{5}\left(\frac{3}{8b}\right)^{3/8} = \frac{8}{5}\left(\frac{c^5}{8N\eta(GM)^{5/3}}\right)^{3/8} = \frac{1}{5}\left(\frac{16}{N^{3/5}}\right)^{5/8}\left(\frac{c^3}{2GM\eta^{3/5}}\right)^{5/8} \tag{A11}$$

Again using $\tau_{\text{S-chirp}}$, we write the integrated orbital phase as

$$\begin{aligned}\Phi(t) &= -\frac{1}{5}\left(\frac{16}{N^{3/5}}\right)^{5/8} \frac{1}{\tau_{\text{S-chirp}}^{5/8}}\left[(t_c - t)^{5/8} - (t_c - t_0)^{5/8}\right] \\ &= -\frac{1}{5}\left(\frac{16}{N^{3/5}}\right)^{5/8}\left[\frac{t_c - t}{\tau_{\text{S-chirp}}}\right]^{5/8} + \Phi_c.\end{aligned} \tag{A12}$$

$\Phi_c = \Phi(t = t_c)$ is called the coalescence phase; it can be chosen to match the data. As we have mentioned previously, the E&M-like model and the linear GR model yield the same waveform (and hence the same integrated orbital phase) as long as the product $\eta^{3/5} N^{3/5} M$ is the same for both models. For the GR model (with $N = 32/5$), we find

$$\Phi(t) = -\left(\frac{2}{5}\right)^{5/8}\left[\frac{t_c - t}{\tau_{\text{S-chirp}}}\right]^{5/8} + \Phi_c. \tag{A13}$$

It is customary to write a single GW observatory's signal from a binary orbit system as

$$h_D(t) = A_D\, a(t) \cos(2\Phi(t) + \Phi_D), \tag{A14}$$

where $A_D$ is a numerical factor that depends on the orbital inclination angle, the polarization orientation angle, and the orientation of the detector's interferometer arms relative to the GW propagation direction. (See Appendix B for a detailed derivation of that factor for vector polarization.) $\Phi_D$ is a phase that depends on the same parameters. We won't need those details in this tutorial. However, they are important in getting information about the polarization of GWs, a critical prediction of general relativity.[43] $a(t)$ is a slowly-varying amplitude, the same for all the observatories, depending on the distance $R$ from the GW source to the detectors, the chirp mass of the binary system, and the orbital frequency:

$$a(t) = N \frac{(GM_{\text{chirp}})^{5/3}}{2Rc^4} \omega^{2/3}(t). \tag{A15}$$

Note that there are various ways to include numerical factors in the definition of $a(t)$. Here we have chosen a combination that involves $M_{\text{chirp}} N^{3/5}$. We have seen previously if that combination has the same value in the E&M-like model and the linear GR model, the GW waveforms will be the same.

Using Kepler's Third Law to replace $\omega(t)$ with $r(t)$ (the binary mass separation) yields

$$a(t) = \frac{N}{8\eta^{1/5}} \frac{r_{\text{S-chirp}}}{R} \frac{r_{\text{S-chirp}}}{r(t)} \tag{A16}$$

Note that Eq. (A16) is the same as Eq. (76) with the Schwarzschild radius associated with the total mass replaced by the Schwarzschild radius associated with $M_{\text{chirp}}$ and the factor of $N$ has been included as explained above.

We now solve Eq. (66) for $r(t)$ in terms of times relative to the coalescence time $t_c$, the time at which the orbital frequency becomes infinite and when the separation between the masses goes to 0. From Eq. (66), and using $r_{\text{S-chirp}}$ in place of $r_S$, we may write

$$\int_{r(t)}^{0} r^3 dr = -\frac{Nc}{4\eta^{1/5}} r_{\text{S-chirp}}^3 \int_{t}^{t_c} d\tau. \tag{A17}$$

After carrying out the integrals and solving for $r(t)$ and using $\tau_{\text{S-chirp}} = r_{\text{S-chirp}}/c$, we obtain

$$r(t) = \frac{N^{1/4}}{\eta^{1/5}} r_{\text{S-chirp}} \left( \frac{t_c - t}{\tau_{\text{S-chirp}}} \right)^{1/4}. \tag{A18}$$

We now use Eq. (A18) in Eq. (A16) to find

$$a(t) = \frac{N}{8\eta^{1/5}} \frac{r_{\text{S-chirp}}}{R} \left( \frac{t_c - t}{\tau_{\text{S-chirp}}} \right)^{-1/4}. \tag{A19}$$

For the GR value $N = 32/5$, we get

$$a(t) = \left( \frac{2}{5} \right)^{3/4} \frac{1}{\eta^{1/5}} \frac{r_{\text{S-chirp}}}{R} \left( \frac{t_c - t}{\tau_{\text{S-chirp}}} \right)^{-1/4}. \tag{A20}$$

We see that the amplitude $a(t)$ increases without limit as $t \to t_c$. Eq. (A20) may be used in Eq. (A14) to express the complete GW waveform for each observatory.

# Appendix B
## Gravitational Wave Detector Response to Vector Gravitational Waves

In this appendix I give the detailed derivation of the GW interferometer detector response to a gravitational wave described as a vector field. For this analysis we will need to specify the orientation of the GW polarization vectors $\hat{e}_1$ and $\hat{e}_2$ relative to the detectors. In practice, with multiple detectors on Earth, it is easiest to denote that orientation relative to Earth as shown in Fig. B1.

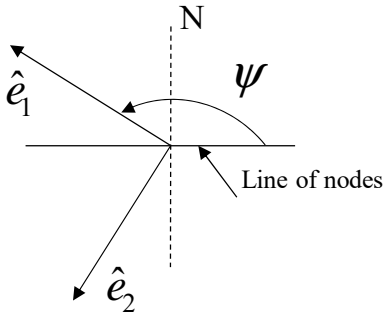

FIG. B1. The definition of the gravitational wave polarization orientation angle $\psi$. The GW propagation direction denoted by $\hat{k}$ (not shown) points up and out of the page. The Line of nodes is the intersection of the wave front $(\hat{e}_1, \hat{e}_2)$ plane and Earth's equatorial plane. The dashed line is perpendicular to Earth's equatorial plane. N indicates Earth's North Pole.

As we saw in the main part of the tutorial, Eq. (49), the gravitational radiation field $\vec{g}_{rad}(t)$ in an E&M-like (vector) model for gravitational waves (GWs) from orbiting binaries can be written as [44]

$$\vec{g}_{rad}(t) = K\left[\hat{e}_1 f_1(\theta_i)\sin 2\Phi(t) + \hat{e}_2 f_2(\theta_i)\cos 2\Phi(t)\right]. \tag{B1}$$

We remind ourselves that $\theta_i$ is the orbital inclination angle relative to the observation direction and $K$ is a parameter that contains the total mass $M = m_a + m_b$, the dimensionless parameter $\eta = m_a m_b / M^2$ ($\eta M$ is the reduced mass of the binary system), the orbital angular frequency $\omega$, the mass separation $r$, and the source-observer distance $R$ dependence:

$$K = \frac{\eta(GM)^{5/3}\omega^{5/3}}{2c^3 R} = \frac{G\eta M \omega^3 r^2}{2c^3 R} \tag{B2}$$

As mentioned previously, $\hat{e}_{1,2}$ are unit vectors perpendicular to the wave propagation direction $\hat{k}$ from the binary orbit to the detector. $\Phi(t)$ is the time-dependent orbital phase of the binaries. (See Appendix A.) The detailed form of $\Phi(t)$ will not be needed in what follows.

We have seen that in the E&M-like model for orbiting binaries[10, 44], the orbital inclination angle dependence is given by

$$\begin{aligned} f_1(\theta_i) &= \sin 2\theta_i \\ f_2(\theta_i) &= -2\sin\theta_i \end{aligned} \tag{B3}$$

In a LIGO-Virgo GW detector interferometer, the beam splitter and mirrors are mounted as pendula, whose oscillation periods and damping times are long compared to the period of the GW oscillations seen in binary black hole and binary neutron star GW events. For each detector, we will denote unit vectors that point along the detector arms as $\hat{d}_x$ and $\hat{d}_y$. In that case, the gravitational radiation field produces a displacement $\Delta x(t)$ of the beam splitter (or a mirror), free to move along the $\hat{d}_x$ direction is given by

$$\Delta x(t) = -\frac{\vec{g}_{rad} \cdot \hat{d}_x}{(2\omega)^2}. \tag{B4}$$

The minus sign tells us that the displacement and acceleration of a harmonic oscillator driven far above its resonant frequency are $\pi$ radians out of phase. The factor of 2 in the denominator reminds us that the GW wave frequency is twice the binary orbital frequency $\omega$, where the orbital frequency is related to the integrated orbital phase by

$$\omega(t) = \frac{d\Phi(t)}{dt}. \tag{B5}$$

The differential movement of the beam splitter ($bs$) and the mirror ($m$) in one arm of the interferometer is proportional to

$$\Delta x_m(t) - \Delta x_{bs}(t) = \frac{K}{4\omega^2} \{\hat{d}_x \cdot \hat{e}_1 f_1(\theta_i)[\sin 2\Phi(t) - \sin 2\Phi(t - \Delta t_x)] \\ + \hat{d}_x \cdot \hat{e}_2 f_2(\theta_i)[\cos 2\Phi(t) - \cos 2\Phi(t - \Delta t_x)]\}, \tag{B6}$$

where

$$\Delta t_x = \frac{L}{c} \hat{k} \cdot \hat{d}_x \tag{B7}$$

is the time it takes the wave to travel the length $L$ of the $x$-arm of the apparatus from the beam splitter to the mirror and $c$ is the usual speed of light. (In what follows, we will assume that the lengths of the two interferometer arms are the same.)

Since $\omega \Delta t_x \ll 1$ for the binary black hole (and the GW170817 binary neutron star) signals detected by the LIGO and Virgo observatories (where the arm length is several km), we may use a Taylor series expansion to express the second term in each of the square brackets in Eq. (B6) as

$$\sin 2\Phi(t - \Delta t_x) \approx \sin 2\Phi(t) - 2\cos 2\Phi(t)\frac{\partial \Phi}{\partial t}\Delta t_x = \sin 2\Phi(t) - 2\omega(t)\Delta t_x \cos 2\Phi(t) \\ \cos 2\Phi(t - \Delta t_x) \approx \cos 2\Phi(t) + 2\sin 2\Phi(t)\frac{\partial \Phi}{\partial t}\Delta t_x = \cos 2\Phi(t) + 2\omega(t)\Delta t_x \sin 2\Phi(t), \tag{B8}$$

where we used Eq. (B5). Deploying Eq. (B8) in the square bracket terms of Eq. (B6), we find that the square bracket terms are approximated by

$$2\omega(t)\Delta t_x \cos 2\Phi(t) \text{ and } -2\omega(t)\Delta t_x \sin 2\Phi(t). \tag{B9}$$

Using Eqs. (B7) and (B9) in Eq. (B6) yields

$$\Delta L_x = \frac{K}{2\omega}\frac{L}{c}\hat{k}\cdot\vec{d}_x[\vec{d}_x\cdot\hat{e}_1 f_1(\theta_i)\cos 2\Phi(t) - \vec{d}_x\cdot\hat{e}_2 f_2(\theta_i)\sin 2\Phi(t)]. \tag{B10}$$

The corresponding analysis for the gravitational-wave-induced change in the length of the y-arm of the interferometer gives

$$\Delta L_y = \frac{K}{2\omega}\frac{L}{c}\hat{k}\cdot\vec{d}_y[\vec{d}_y\cdot\hat{e}_1 f_1(\theta_i)\cos 2\Phi(t) - \vec{d}_y\cdot\hat{e}_2 f_2(\theta_i)\sin 2\Phi(t))]. \tag{B11}$$

The LIGO-Virgo gravitational wave signal $h(t)$ is the differential strain $\Delta L/L$ between the two interferometer arms:

$$h(t) = \frac{\Delta L_x - \Delta L_y}{L} = \frac{G\eta M \omega^2 r^2}{4c^4 R}[f_1(\theta_i)\cos 2\Phi(t)\{(\hat{k}\cdot\vec{d}_x)(\vec{d}_x\cdot\hat{e}_1) - (\hat{k}\cdot\vec{d}_y)(\vec{d}_y\cdot\hat{e}_1)\}$$
$$+ f_2(\theta_i)\sin 2\Phi(t)\{(\hat{k}\cdot\vec{d}_y)(\vec{d}_y\cdot\hat{e}_2) - (\vec{w}_z\cdot\vec{d}_x)(\vec{d}_x\cdot\hat{e}_2)\}]. \tag{B12}$$

The result in square brackets in Eq. (B12) agrees with those of Ref. [38, 45] (Eqs. (26) and (27)) and Ref. [4] (Eqs.(14)-(15)) except for an overall minus sign difference in the $f_2(\theta_i)$ term. That sign difference is not important in comparing signal amplitudes and phase differences among the three LIGO-Virgo detectors.

The terms inside the braces in Eq. (B12) are called the antenna pattern functions for the so-called x-polarization and y-polarization:

$$F_x(\hat{w},\hat{d}) = (\hat{k}\cdot\vec{d}_x)(\vec{d}_x\cdot\hat{e}_1) - (\hat{k}\cdot\vec{d}_y)(\vec{d}_y\cdot\hat{e}_1)$$
$$F_y(\hat{w},\hat{d}) = (\hat{k}\cdot\vec{d}_y)(\vec{d}_y\cdot\hat{e}_1) - (\hat{k}\cdot\vec{d}_x)(\vec{d}_x\cdot\hat{e}_1), \tag{B13}$$

which depend on the GW vector $\hat{w} = (\hat{e}_1,\hat{e}_2,\hat{k})$ and the unit vectors giving the interferometer arm orientations. ($\hat{k}$ is the same as the unit vector $\hat{R}$, which indicates the direction from the source to the observation point.) It is rather remarkable that this vector field calculation gives exactly the same antenna patterns as those based on general metric theories of gravity.[38] The lesson is that the polarization properties are primarily geometric quantities (scalar, vector, or tensor), independent of the underlying dynamical theory.

By way of comparison, I note that in Einstein gravity, GWs have two "tensor" polarization modes called "plus" and "cross." The corresponding GW signal turns out to be [38, 45]

$$h(t) = K[f_+(\theta_i) F_+(\hat{w},\hat{d}) \cos 2\Phi(t) + f_\times(\theta_i) F_\times(\hat{w},\hat{d}) \sin 2\Phi(t)], \tag{B14}$$

where

$$F_+(\hat{w},\hat{d}) = \tfrac{1}{2}\left[ (\hat{e}_1 \cdot \hat{d}_x)^2 - (\hat{e}_1 \cdot \hat{d}_y)^2 - (\hat{e}_2 \cdot \hat{d}_x)^2 + (\hat{e}_2 \cdot \hat{d}_y)^2 \right]$$
$$F_\times(\hat{w},\hat{d}) = (\hat{e}_1 \cdot \hat{d}_x)(\hat{e}_2 \cdot \hat{d}_x) - (\hat{e}_1 \cdot \hat{d}_y)(\hat{e}_2 \cdot \hat{d}_y) \tag{B15}$$

and

$$f_+(\theta_i) = (1+\cos^2\theta_i)/2$$
$$f_\times(\theta_i) = -\cos\theta_i . \tag{B16}$$

You can see that the strain signal depends on the angle of inclination as well as on the orientation of the GW propagation and polarization directions relative to the detector unit vectors for both the GR tensor polarization modes and the E&M-like vector modes. It is important to note that the detector responses are different for the two types of polarization, so observations that are sensitive to GW polarization provide a unique test of general relativity.